\documentclass[1p]{elsarticle}
\usepackage{graphicx,psfrag}
\usepackage{subfigure}
\usepackage{amsmath,amssymb}
\usepackage{type1cm}
\usepackage{bm}
\usepackage{comment}
\usepackage{mathptmx}
\usepackage{wrapfig}

\newcommand\beq{\begin{equation}}
\newcommand\eeq{\end{equation}}
\newcommand\beqa{\begin{eqnarray}}
\newcommand\eeqa{\end{eqnarray}}
\renewcommand{\d}{\partial}
\newcommand{\ds}[1]{#1 \hspace{-0.5em}/}  
\newcommand\bzeta{\mbox{\boldmath$\zeta$}}

\newcommand\bk{{\bf k}}
\newcommand\bq{{\bf q}}
\newcommand\E{\epsilon}

\newcommand\bp{{\bf p}}

\begin{document}
\title{Spontaneous magnetization in QCD and non-Fermi-liquid effects}

\author{K. Sato\corref{cor1}}
\ead{k-sato@ruby.scphys.kyoto-u.ac.jp}

\author{T. Tatsumi}

\address{ Department of Physics, Kyoto University, Kyoto 606-8502, Japan}
\cortext[cor1]{Corresponding author}

\begin{abstract}
Magnetic properties of quark matter at finite temperature are discussed by evaluating the
 magnetic susceptibility. 
Combining the microscopic calculation of the self-energy for quarks as
 well as the screening effects for gluons with Fermi-liquid theory in a consistent way, 
we study the temperature dependence of the
 magnetic susceptibility. 
The longitudinal gluons have the static screening given by the Debye
 mass, and 
 have a standard temperature dependence of $O(T^2)$. 
An anomalous $T^2\ln T$ term arises in the magnetic susceptibility 
as a novel non-Fermi-liquid effect 
due to the anomalous self-energy for quarks given by the 
dynamic screening for transverse gluons. 
We then extract the critical (Curie)
 temperature and present the magnetic  
phase diagram on the density-temperature plane. 
\end{abstract}


\maketitle

\section{Introduction}

Nowadays the phase diagram of QCD has been extensively studied 
theoretically by lattice calculations \cite{R-lat} and the effective models
or experimentally by high-energy heavy-ion collisions (RHIC,LHC) \cite{R-ion} 
and observation of 
compact stars \cite{R-star}. In particular, matter at high density but not
high 
temperature should be interesting, since the Fermi surface is a good concept there 
and we can study interesting particle correlations such as
superconducting pairing or 
instability of the Fermi surface in 
a clear way. It is also related with a quest of "new form of matter" 
inside compact stars or during their thermal evolution \cite{R-star2}.  

We discuss here magnetic properties of QCD by the use of
Fermi-liquid theory \cite{bay04,noz,bay76}. A possibility of spontaneous spin polarization has
been suggested by one of the authors \cite{tat00,nak03}, using the
one-gluon-exchange (OGE) interaction. Higher-order calculations for the
free energy have
also supported this idea \cite{nak03, nie05}.
If it is realized,
it may give a microscopic origin of strong magnetic field observed in
compact stars, especially magnetars \cite{mag}. The mechanism of spontaneous magnetization is very much 
similar to the one for the electron gas \cite{blo29,her66,mor85}, where the leading-order
contribution 
comes from the Fock exchange interaction. Electrons with the same spin
can 
effectively avoid the repulsive Coulomb interaction due to the Pauli
principle. On the other hand, the kinetic energy is increased as the number of each
spin is different. Hence spontaneous polarization may occur at a peculiar
density.    
Differences are color and flavor degrees of freedom for quark matter.

We apply the Landau Fermi-liquid theory (FLT) to elucidate the
critical behavior of the magnetic phase transition at finite density
and temperature. We evaluate the magnetization of quark matter by applying 
a small magnetic field $\bf B$. Then the magnetic susceptibility appears as a
coefficient of the linear term in $B$. The divergence and sign change of the magnetic
susceptibility signal the magnetic instability to the ferromagnetic
phase, since its inverse measures the curvature of the free energy at
the origin with
respect to the magnetization. 
Thus quarks near the Fermi surface are
responsible to 
the magnetic transition and the spin dependent quark-quark interaction and the density
of states near the Fermi
surface are the key ingredients within FLT
\footnote{The high-density effective theory should be also a powerful
tool in this context \cite{hon,sch}, and its relation to FLT
has been discussed \cite{han04}.}
.   

In this theory, quarks are treated as quasi-particles and their
energy $E$ is regarded as a functional of the distribution
function. FLT is restricted to the low-lying excitations around the
Fermi surface , where the lifetime of quasi-particles are long enough
to be regarded as ``particles'' with properties resembling those of free
particles.
The single-particle energy is then given by the variation of $E$
with respect to the distribution function, incorporating the 
self-energy; the quasi-particle interactions are given by the variation
of the single-particle energy with respect to the distribution function 
and they are specified by the set of a few parameters, called the
Landau-Migdal parameters. 
It is well-known that there appear infrared (IR) singularities 
in the Landau-Migdal parameters in gauge theories (QED/QCD) since the
gauge interaction is infinite range. 
To improve the IR behavior we must take into account the correlation
effects for 
the quasi-particle interaction: we must sum up an infinite series of
the most divergent diagrams as the self-energy term of the gauge field.
Thus we take into account the screening effects for
the gauge field. 
For example, the Coulomb interaction becomes ``finite'' range and there is left no
singularity in the condensed-matter physics \cite{noz}. 
The summation of the infinite series or the inclusion of the screening effects is also required
by the argument of the hard-dense-loop (HDL) resummation \cite{kap}. 
Consider the particle-hole polarization diagram as a leading-order
contribution to the self-energy of the gauge field. 
Since the IR singularities
appear for quasi-particles with the collinear momenta, the soft
gluon should give a dominant contribution. Then the particle-hole polarization function
$\Pi(q)\sim O(g^2 \mu ^2)$ should be the same order of magnitude with the
energy-momentum squared of gluons $\sim (g \mu)^2 $.  Hence we must sum
up the infinite series, $1/q^2\left(1+\Pi/q^2+...\right)$, to get the 
meaningful results. For the longitudinal mode we can 
see the static screening by the Debye
mass and the IR behavior is surely improved. On the other hand, there is no
static screening for the transverse mode but only the dynamic 
screening due to the Landau damping\cite{kap}. 
Thus the IR singularities  still remain in the Landau-Migdal parameters
derived 
through the exchange of the gauge boson of the transverse mode.

In a recent paper we studied the screening effects for gluons 
to evaluate the magnetic susceptibility of quark
matter at $T=0$ within 
FLT 
\cite{tat08, tat082}. We have seen that the
transverse gluons 
still give logarithmic singularities to the Landau-Migdal parameters,
but they cancel each other 
in the magnetic susceptibility to give a finite result. Finally the static screening
for the longitudinal gluons gives the $g^4{\rm ln}(1/g^2)$ term in the
magnetic susceptibility \cite{her66,bru57}. A similar term is also
obtained as a correlation effect in electron gas and the ferromagnetic
region is diminished. However,  
we also find that this term has an interesting behavior in QCD, depending on 
the number of flavors. Consequently the screening effect does not 
necessarily work against the magnetic instability, which is a different
aspect from electron gas.

In this paper, we extend our previous analysis to the finite-temperature
case and figure out the non-Fermi-liquid aspect of the magnetic
susceptibility. Since thermal gluons should give higher-order terms in $T$, we
only consider the interacting quark system within FLT. 
We take into account the effect of the self-energy of
quasi-particles by using the explicit expression given by the
perturbation method. We shall see that the longitudinal gluons give
little temperature dependence due to the static screening,    
while the transverse gluons exhibit an
anomalous temperature dependence for the magnetic susceptibility through 
the dynamic screening.  
Shanker has shown that the application of the renormalization group (RG) 
automatically leads to FLT as  
a fixed-point, where all the quasi-particle 
interactions are {\it marginal}
\cite{sha94}, except the attractive superconducting (BCS) channel. 
This argument, however,
 may be applied
to the case of the short-range interaction. 
Recent RG arguments have revealed 
that the quasi-particle interaction through the exchange of the
transverse gauge bosons is {\it relevant} and induces the non-Fermi-liquid
effects in gauge theories (QED/QCD) \cite{sch,boy01,gan,cha}. 
Furthermore, it should be important to note that quasi-particle interaction
is {\it infrared free},  so that the effective coupling becomes very weak when the
excitation energy of the quasi-particles from the Fermi surface
approaches 
to null \cite{sch}. The effective coupling constant is given by the product of the gauge
coupling constant $g^2$ and the Fermi velocity $v_F$, $C_{\rm eff}=g^2
v_F/4\pi$. Since $v_F$ goes to zero on the Fermi surface, 
the perturbation method should be meaningful even if the
gauge coupling is not weak. 

The quasi-particle energy $\epsilon_\bk$ should be given by the
root of the transcendental equation, $\epsilon_\bk=E_\bk+{\rm Re}\Sigma_+(\epsilon_\bk, \bk)$
with $E_\bk=(|\bk|^2+m^2)^{1/2}$, 
and the one-loop result of the self-energy $\Sigma_+(\E_\bk, \bk)$ by
way of the microscopic many-body technique   
exhibits an anomalous behavior as $\E_\bk\rightarrow \mu $, 
${\rm Re}\Sigma_+(\E_\bk)\sim \gamma(\E_\bk-\mu ){\rm ln}(\Lambda/|\E_\bk- \mu |)$
with $\gamma=g^2/9\pi^2$, due to the
dynamic screening \cite{boy01,man}. 
The further analysis of the self-energy by
way of the Schwinger-Dyson equation or the renormalization group have
shown that the anomalous term in the one-loop result is reliable even if 
$\gamma\ln(\Lambda/|\E_\bk-\mu |) \sim 1$: actually there is no difference
between the one-loop result and the solution of the Schwinger-Dyson
equation \cite{sch}.
Consequently, the renormalization factor 
$z_+(k)=(1-\partial{\rm Re}\Sigma_+(\omega)/\partial\omega)^{-1}|_{\omega=\E_\bk}$
vanishes at $k=k_F$, which means that 
the distribution function has no discontinuity at $\epsilon_{k_F}=\mu $
even at $T=0$, or there is {\it no Fermi surface} \cite{noz,lut,lut61}. 
However, the distribution function of the quasi-particles is still the
step function at $T=0$. Actually we have seen that 
the transverse mode causes no effect for the instability of the Fermi surface due to its
sharpness \cite{tat08}. 

At $T\neq 0$, the Fermi surface is diffused over the width of $O(T)$ , so 
that the transverse gluons should give rise to an anomalous logarithmic dependence 
for physical quantities through the dynamic screening, as we shall see in the following.   
For instance, the anomaly in the specific heat has been
repeatedly discussed in the literatures
\cite{boy01,hol,rei,gan,cha,ipp,ger05}; 
the transverse gauge
field gives a $T\ln T$ term as a leading-order contribution. 
Analogous non-Fermi-liquid behavior can be also seen in the context of
color superconductivity at high densities \cite{son,sch99,bro,wan}.
We shall see
that there appears a $T^2\ln T$ term in the magnetic susceptibility as
another non-Fermi-liquid effect \cite{tat083}. This temperature
dependence is different from that of the specific heat, while one may
naively expect the same one by considering the density of states on the
Fermi surface: the leading-order contribution is $\ln T$,
but it is canceled by another $\ln T$ term coming from the
spin-dependent Landau-Migdal parameter to leave the next-to-leading-order term
of $T^2\ln T$ in the magnetic susceptibility.

To extract the temperature dependence correctly we must consider the temperature
dependence of the chemical potential as well. We derive it within FLT
and show that the result is the same as that given by a microscopic
calculation \cite{ipp}.      
Properly taking into account the temperature dependence of the chemical
potential, which also includes the $T^2\ln T$ term, we finally present 
the magnetic susceptibility at finite temperature. 

If the magnetic
susceptibility  diverges at
some density and some temperature, it signals the magnetic instability
of quark matter to the ferromagnetic phase. Thus we can extract the
critical densities and temperatures to draw a critical
curve on the density-temperature plane.  
Using our results, we shall demonstrate
the magnetic phase diagram, which might be
useful in considering some phenomenological implications of 
magnetism of quark matter. We get several tens MeV as the maximum
critical (Curie) temperature. This value is similar to that expected 
during supernova explosion, so that 
it would be interesting to consider the
thermal evolution of magnetars in the initial cooling stage. 
  
In this work we sometimes need the explicit result of a microscopic
calculation rather than general framework of FLT.  
In each step of the calculation we try to figure out the difference and
similarity, compared with the usual discussion of FLT.

In \S 2 a framework is presented within Fermi-liquid theory to
study the magnetic properties of quark matter. 
In \S 3 non-Fermi-liquid effects are 
extracted in the magnetic susceptibility  at finite temperature. Some
specific temperature dependence of chemical potential is also discussed.
In \S 4 a magnetic phase diagram is presented on the density-temperature
plane, where we can also see an importance of the non-Fermi-liquid effect.    
Some relations between the microscopic calculation and the FLT are
summarized in Appendix B.

\section{Fermi-liquid theory for quark matter}

Within the Landau Fermi-liquid theory (FLT) we assume a one-to-one correspondence
between the states of the free Fermi gas and those of the interacting
system. Quarks are treated as quasi-particles carrying the same quantum
numbers of the free particles, and the quasi-particle distribution function is simply
given by the Fermi-Dirac one,
\beq
n(\bk,\zeta)=\left[1+\exp(\beta(\epsilon_{\bk,\zeta}-\mu))\right]^{-1}
\eeq
with the quasi-particle energy $\epsilon_{\bk,\zeta}$ specified by the
momentum $\bk$ and a spin quantum number $\zeta=\pm1$. 

Since ``spin'' is not a good quantum number in relativistic theories, we
specify the polarization of each quark by introducing the spin vector
$a^\mu$, which is a covariant generalization of the spin direction
$\bm{\zeta}$ (we take it along $z-$axis) in the 
non-relativistic case. Actually spin vector $a^\mu$ is a space-like four vector with the
constraints, $a^\mu k_\mu=0$ and $a^2=-1$, and it is 
reduced to the three vector $\bm{\zeta}$ in the rest frame. It is not
uniquely given, but  
there are many choices for the explicit
form of spin vector. We must choose the most relevant one for the
description of ferromagnetic phase\cite{tat00}. Here we assume the 
standard form \cite{itz,ber},
\beq
{\bf a}=\bm{\zeta}+\frac{\bk(\bm{ \zeta} \cdot \bk)}{m(E_k+m)}, ~~~a^0=\frac{\bk\cdot\bm{\zeta}}{m},
\eeq
with $E_k=(\bk^2+m^2)^{1/2}$.

We can define an eigenstate $u^{(\zeta)}$ of the operator,
$W\cdot a$, where $W^\mu$ is the Pauli-Lubanski vector, 
$W_\mu=-1/4\epsilon_{\mu\nu\rho\sigma}k^\nu\sigma^{\rho\sigma}$:
\beq
-\frac{W\cdot a}{m}u^{(\zeta)}=\frac{\zeta}{2}u^{(\zeta)},
\eeq
with $\zeta=\pm 1$. Accordingly we can define polarization density
matrix $\rho(k,\zeta)$,
\beq
\rho(k,\zeta)=\frac{1}{2m}(\ds{k}+m)P(a),
\eeq
by the projection operator, $P(a)=1/2\cdot(1+\gamma_5\ds{a})$.

\subsection{Quasi-particle interaction}

Since the color-singlet magnetization is not directly related to the
color degree of freedom, we,
hereafter,  only consider the color-symmetric interaction among
quasi-particles that can be written as the sum of two parts, 
the spin symmetric ($f^s_{\bk,\bq}$) and
anti-symmetric ($f^a_{\bk,\bq}$) terms;
\beq 
f_{\bk\zeta,\bq\zeta'}=f^s_{\bk,\bq}+\zeta\zeta'f^a_{\bk,\bq}.
\eeq
Since quark matter is color singlet as a whole, the Fock exchange
interaction gives a leading contribution \cite{ohn}. We, hereafter, take
a perturbation technique and 
consider the one-gluon-exchange interaction (OGE) with relatively large 
coupling constant. One may wonder the perturbation method should break in
that case. However, the {\it real} expansion parameter is not
the gauge coupling constant in the present case; RG argument has shown
that the quasi-particle interaction
is infrared free and the effective coupling becomes very weak
 for the low-lying excitations around the Fermi surface\cite{sch}. 
The effective coupling constant is then given by the product of the gauge
coupling constant $g^2$ and the Fermi velocity $v_F$, $C_{\rm eff}=g^2
v_F/4\pi$, and $v_F$ becomes vanished on the Fermi surface. 
Hence the perturbation method is meaningful even if the
gauge coupling is not weak. 

For a pair with color
index $(a,b)$, the Fock exchange interaction gives a factor 
$(\lambda_\alpha)_{ab}(\lambda_\alpha)_{ba}=1/2-1/(2N_c)\delta_{ab}$, which 
is always positive for any pair. Hence the situation is very similar to
electron gas in QED. Since we are interested in the electromagnetic
properties of quark matter, only the color symmetric interaction is 
relevant, which is written as 
\beq
f_{\bk\zeta,\bq\zeta'}
=\frac{1}{N_c^2}\sum_{a,b}f_{\bk\zeta
a ,\bq\zeta' b }
=\frac{m}{E_k}\frac{m}{E_q}M_{\bk\zeta,\bq\zeta'},
\eeq
with the invariant matrix element,
\beq
M_{\bk\zeta,\bq\zeta'}=-g^2\frac{1}{N_c^2}{\rm
tr}\left(\lambda_\alpha/2\lambda_\alpha/2\right)M^{\mu\nu}(k,\zeta; q,\zeta')D_{\mu\nu}(k-q),
\eeq
where $M^{\mu\nu}(k,\zeta; q,\zeta')
={\rm
tr}\left[\gamma^\mu\rho(k,\zeta)\gamma^\nu\rho(q,\zeta')\right]$. The
matrix elements are easily evaluated with the definition of the density
matrix $\rho(k,\zeta)$ (Eq.~(4)) and summarized in Appendix A.
Since the OGE interaction is a long-range force and we
consider the small energy-momentum transfer between quasi-particles, we
must treat the gluon propagator by taking into account HDL 
resummation \cite{kap}. Thus we take into account the screening effect,
\beq
D_{\mu\nu}(k-q)=P^t_{\mu\nu}D_t(p)+P^l_{\mu\nu}D_l(p)-\xi\frac{p_\mu
p_\nu}{p^4} 
\eeq
with $p=k-q$, where $D_{t(l)}(p)=(p^2-\Pi_{t(l)})^{-1}$, and 
the last term represents the gauge dependence with a parameter
$\xi$. $P^{t(l)}_{\mu\nu}$ is the projection operator onto the
transverse (longitudinal) mode, 
\beqa
P^t_{\mu\nu}&=&(1-g_{\mu 0})(1-g_{\nu
0})\left(-g_{\mu\nu}-\frac{p_\mu p_\nu}{|\bp|^2}\right)\nonumber\\
P^l_{\mu\nu}&=&-g_{\mu\nu}+\frac{p_\mu
p_\nu}{p^2}-P^t_{\mu\nu}.
\eeqa
The self-energies for the transverse and longitudinal gluons are given
as 
\beqa
\Pi_l(p_0,{\bf p})&=&\sum_{f=u,d,s} \left( m_{D,f}^2+i\frac{\pi
m_{D,f}^2}{2u_{F,f}}\frac{p_0}{|\bp|} \right)\nonumber\\
\Pi_t(p_0,{\bf p})&=&-i\sum_{f=u,d,s} \frac{\pi u_{F,f} m_{D,f}^2}{4}\frac{p_0}{|\bp|}, 
\eeqa
in the limit $p_0/|\bp|\rightarrow 0$, with $u_{F,f}\equiv
k_{F,f}/E_{F,f}$ and the Debye mass for each flavor,
$m_{D,f}^2\equiv g^2\mu_f k_{F,f}/2\pi^2$ \cite{kap}
\footnote{The Debye mass is given as $e^2\mu^2u_F/\pi^2$ for 
electron gas in QED.}
. 
Thus the longitudinal gluons are statically
screened to have the Debye mass, while the transverse gluons are dynamically screened by the Landau damping,
in the limit $p_0/|{\bf p}|\rightarrow 0$. Accordingly, the screening effect for the transverse gluons is ineffective 
at $T=0$, where soft gluons ($p_0/|\bp|\rightarrow 0$) contribute. 
At finite temperature, gluons with $p_0 \sim O(T)$ can contribute due to the diffuseness of the Fermi surface and the transverse gluons
are effectively screened.

\subsection{Magnetic susceptibility}
We consider the linear response of the normal(unpolarized) quark matter
by applying a small magnetic field 
$\bf B$. 
The magnetic susceptibility is defined as
\beq
\chi_M=\sum_{f=u,d,s}\chi_M^f=\sum_{f=u,d,s}
\left.\frac{\partial\langle M\rangle_f}{\partial B}\right|_{B=0}
\eeq
with the magnetization $\langle M\rangle_f$ for each flavor, 
where we take ${\bf B}//\hat {\bf z}$.  
We consider here the spin susceptibility, so that the magnetization is
given by $\langle M\rangle_f=V^{-1}\langle\int d^3x{\bar q}_f\Sigma_z
q_f\rangle$ with $\mu _q^f$ being the Dirac magnetic moment.
Hereafter, we shall concentrate on one flavor and omit the flavor indices 
because the magnetic susceptibility is given by the sum of the contribution from each flavors.
Magnetic susceptibility is then written in terms of the quasi-particle 
interaction~\cite{bay04,tat082},
\beqa
\chi_M=\left(\frac{\bar g_D\mu_q}{2}\right)^2\frac{N(T)}{1+N(T)\bar
f^a}
\eeqa
where $\bar g_D=\int_{|\bk|=k_F}d\Omega_\bk/4\pi g_D(\bk)$ is the gyromagnetic ratio
, 
\beq
g_D(\bk)\zeta=2{\rm
tr}\left[\Sigma_z\rho(k,\zeta)\right]=2
\left(a_z-\frac{k_z}{E_k}a_0\right),
\eeq
and can be explicitly written as 
\beq
g_D(\bk)=2\left[1-\frac{k_z^2}{E_k(E_k+m)}\right],
\eeq
using the spin vector (Eq.~(2)).

$N(T)$ is the effective density of states on the Fermi surface, and 
defined by 
\beq
N(T)=-\int_{-\infty}^\infty d\omega\frac{\partial n(\omega)}{\partial\omega}\nu(\omega),
\label{dos}
\eeq
where $n(\omega)$ is the Fermi-Dirac distribution function and 
$\nu(\omega)$ is the density of states of the quasi-particles at
energy $\omega$,
\beq
\nu(\omega)\equiv -\frac{2N_c}{\pi}\int\frac{d^3k}{(2\pi)^3}
{\rm Im}\left[{\cal S}_+^R(\omega, \bk)\right],
\label{eq:nu}
\eeq
in terms of the retarded Green function of the quasi-particles, 
${\cal S}_+^R(\omega, \bk)=(\omega-\epsilon_\bk-i\eta)^{-1}$.
The quasi-particle energy $\epsilon_\bk$ is given by the kinetic
energy $E_\bk=\sqrt{\bk^2+m^2}$ and the self-energy
$\Sigma_+(\epsilon_\bk,\bk)$,  
\beq
\epsilon_\bk=E_\bk+\Sigma_+(\epsilon_\bk,\bk). \label{eq:epsilon}
\eeq
Since $n(\omega) \rightarrow \theta(\mu -\omega)$ as $T\rightarrow
0$, $N(T)$ is reduced to the effective density of states at the Fermi surface
of the quasi-particles at $T=0$, $N(0)=N_ck_F^2/\pi^2 v_F$ with the
Fermi velocity, $v_F=(\partial\epsilon_\bk/\partial k)_{k=k_F}$. 
Note that $-\partial n(\omega)/\partial\omega$ is
sharply peaked at $\omega=\mu $ for $T/\mu \ll 1$ which we are concerned
with, and we can see that only the
quasi-particles with the energy $\epsilon_\bk\simeq\mu$ still gives a dominant
contribution
\footnote{We shall see that the critical (Curie) temperature $T_c$ is
actually 
small compared with mass or Fermi momentum such that $T_c/m$ or
$T_c/k_F\ll 1$.}
.


$\bar f^a$ is a Landau-Migdal parameter defined by 
\beq
{\bar f^a}\equiv -2N_c\int\frac{d^3k}{(2\pi)^3}\frac{\partial
n(\epsilon_\bk)}{\partial\epsilon_\bk}{\bar f}^a_{k,k_s}/N(T)
\eeq
after angle-averaged over the Fermi surface,
\beq
{\bar f^a}_{k,k_s}\equiv \int\frac{d\Omega_\bk}{4\pi}\int\frac{d\Omega_\bq}
{4\pi}\left.f^a_{\bk,\bq}\right|_{|\bq|=k_s},
\eeq
where $k_s$ is defined by $\epsilon_{k_s}=\mu $ and coincides with the usual Fermi momentum $k_F$ at $T=0$. 

\section{Non-Fermi-liquid behavior at finite temperature}

We consider the magnetic
susceptibility at low temperature. In the usual FLT it should have little temperature
dependence, while we shall see that the anomalous $T^2\ln T$ term
appears due to the dynamic screening effect; the
anomalous term is a leading-order contribution at low temperature. To
show such term we carefully analyze the self-energy $\Sigma_+(\E_\bk)$, 
the derivative of which logarithmically diverges as $\E_\bk \rightarrow
\mu$. 

\subsection{Average of the density of states near $\epsilon_\bk=\mu $}

In this subsection, we study the average of the density of state given by Eq.~(\ref{dos}).
As we shall see, we can not use the usual low-temperature expansion
because of the singularity of the quasi-particle 
energy at the Fermi surface. Therefore, more careful treatment is needed. 

First, we substitute Eq. (\ref{eq:nu}) into Eq. (\ref{dos}) and change the variable of integration in Eq. (\ref{dos})
from $k$ to $\E_\bk(=\omega)$;
\beq
N(T)=\frac{N_c}{\pi^2} \int_{\E_0}^{\infty} d\omega  
\frac{d k}{d \omega} k^2 \frac{\beta e^{\beta\left(\omega-\mu \right)}}{\left(e^{\beta\left(\omega-\mu \right)}+1\right)^2}  , \label{eq:3.1.1}
\eeq
with $\E_0 \equiv \E_{|\bk|=0}$. The quasi-particle energy $\omega$
satisfies the relation,  
\beq
\omega = E_{k(\omega)} +{\rm Re}\Sigma_+(\omega,k(\omega)), \label{eq:omega}
\eeq 
where we discard the imaginary part within the quasi-particle
approximation. 

The one-loop self-energy is almost independent of the momentum, and can be written
 as \cite{boy01,man}
\begin{align}
{\rm Re}\Sigma_+(\omega,k)\sim
{\rm Re} \Sigma_+(\mu ,k_F)&-\frac{C_fg^2u_{F}}{12\pi^2}(\omega-\mu )\ln\frac{\Lambda}{|\omega-\mu |} \notag\\
&+\Delta^{\rm reg}(\omega-\mu)
\label{eq:Sigma}
\end{align} 
and 
\beq
{\rm Im}\Sigma_+(\omega,k) \sim \frac{C_fg^2}{24\pi}|\omega-\mu |
+O\left((\omega-\mu )^2\right)
\label{eq:ImSigma}
\eeq 
around $\omega \sim \mu $ with $C_f=(N_c^2-1)/(2N_c)$ and $u_F=k_F/E_{k_F}$. $\Lambda$ is a
cut-off factor and should be an order of the Debye mass. (In this paper, we take $\Lambda=M_D=\sqrt{\sum_f m_{D,f}^2}$.) Note that the
 anomalous term in Eq.~(\ref{eq:Sigma}) appears 
from the dynamic screening of the transverse gluons, and the
 contribution by the longitudinal gluons is summarized  
in the regular function $\Delta^{\rm reg}(\omega-\mu)$ of $O(g^2)$. 
Within the approximation given by Eqs. (\ref{eq:Sigma}) and (\ref{eq:ImSigma}), the self-energy is independent
of spatial momentum $k$  and thus we omit the argument $k$
 hereafter. The renormalization factor $z_+(k)$ is then given by the
 equation,
 $z_+(k)=(1-\partial{\rm Re}\Sigma_+(\omega)/\partial\omega|_{\omega=\epsilon_\bk})^{-1}$, and we have 
\beq
z_+(k)^{-1}\sim -\frac{C_fg^2u_F}{12\pi^2}\ln|\epsilon_\bk-\mu|.
\eeq
It exhibits a logarithmic divergence as $\epsilon_\bk\rightarrow \mu$,
 which causes non-Fermi-liquid behavior \cite{sch}.

Differentiating Eq. (\ref{eq:omega}) with respect to $k$, we find 
\begin{align}
\frac{dk(\omega)}{d\omega} \simeq\left(1- \frac{\partial {\rm Re}\Sigma_+(\omega)}{\partial \omega}\right)
\frac{E_{k(\omega)}}{k(\omega)}.
\label{jac}
\end{align}
Here, we have used $\partial {\rm Re} \Sigma_+/\partial k \simeq 0$ for the self-energy Eq. (\ref{eq:Sigma}). 

Eventually, $N(T)$ is written as,
\begin{align}
N(T) \simeq \frac{N_c}{\pi^2} \int_{\E_0}^{\infty} d\omega \left(1-
 \frac{\partial{\rm Re} \Sigma_+(\omega)}{\partial
 \omega}\right)k(\omega)E_{k(\omega)} 
\frac{\beta e^{\beta\left(\omega-\mu \right)}}{\left(e^{\beta\left(\omega-\mu \right)}+1\right)^2} .\label{eq:NTomega}
\end{align}
As is discussed in Appendix B, we can separate the contribution by the
 longitudinal gluons $N_l(T)$ from $N(T)$. Since the longitudinal gluon
 exchange is short-ranged by the Debye screening mass, it becomes
 almost temperature independent,   
\beq
N_l(T)=\frac{N_ck_FE_F}{3\pi^2}f_{l;1}^s,
\label{NTl}
\eeq
with the Landau-Migdal parameter $f_{l;1}^s$,
\begin{eqnarray}
f_{l;1}^s=-\frac{3N_c^{-1}C_fg^2}{8E_F^2k_F^2}\left[\kappa k_F^2+2E_F^2\right]
\left[(1+\kappa)I_0(\kappa)-1\right],
\label{f1ls}
\end{eqnarray}
where $\kappa=\sum_f m_{D,f}^2/2k_F^2$
and
\beqa
I_0(\kappa)&=&\frac{1}{2}\int_{-1}^1\frac{du}{1-u+\kappa}
\simeq \frac{1}{2}{\rm
      ln}\left(\frac{2}{\kappa}\right)
\simeq {\rm ln}(g^{-2}).
\eeqa

To evaluate the transverse contribution, $N_t(T)=N(T)-N_l(T)$, we only use
 the transverse part in Eq.~(\ref{eq:Sigma}):
substituting Eq. (\ref{eq:Sigma}) into Eq. (\ref{eq:NTomega}), we obtain 
the leading-order contribution
\footnote{We discard here the temperature independent term of $O(g^2)$,
 which cannot be given only by Eq. (\ref{eq:Sigma}). However, we can
 recover it by taking the $T\rightarrow 0$ limit later (see subsection 3.3). 
}
,
\begin{multline}
N_t(T)=\frac{N_ck_s\mu }{\pi^2}\Big[
 1+\frac{\pi^2}{6}\frac{(2k_F^2-m^2)}{k_F^4}T^2 \\
+\frac{C_fg^2u_F}{24}\frac{(2k_F^2-m^2)}{k_F^4}
T^2\ln\left(\frac{\Lambda}{T}\right)
 +\frac{C_fg^2u_F}{12\pi^2}\ln\left(\frac{\Lambda}{T}\right)
 \Big]+O(g^2T^2)
,\label{eq:NT1}
\end{multline} 
after some manipulation (Appendix C).
$N_t(T)$ has a term proportional to $\ln T$, which gives a singularity at
 $T=0$. This singularity corresponds to 
the logarithmic divergence of the Landau-Migdal parameter $f_1^s$ at
 $T=0$ (Appendix B).
Inverting this for $T\neq 0$, we find
\begin{multline}
N^{-1}(T)=\frac{\pi^2}{N_ck_s\mu }\Big[
 1-\frac{\pi^2}{6}\frac{(2k_F^2-m^2)}{k_F^4}T^2 \\
-\frac{C_fg^2u_F}{72}\frac{(2k_F^2-m^2)}{k_F^4}
T^2\ln\left(\frac{\Lambda}{T}\right) -\frac{C_fg^2u_F}{12\pi^2}\ln\left(\frac{\Lambda}{T}\right) \Big]
 -N_l(T)+O(g^2T^2). \label{eq:NTinv1}
\end{multline}
Note that this formula is meaningful only at nonzero temperature; a
 renormalization group argument tells that theory is infrared free in
 this case and the perturbation analysis of the low energy behavior is
 reliable \cite{sch}. We have kept the next-to-leading term ($T^2\ln T$
 term) as well as the leading-order term ($\ln T$ term), since 
we shall see that the $\ln T$ term is canceled out
 by another term appearing 
in the spin-dependent Landau-Migdal parameter $\bar f^a$ in the magnetic
 susceptibility. 

\subsection{Cancellation of the singular terms at $T=0$}

At finite temperature, the magnetic susceptibility is given by
\begin{equation}
\chi_M=\left(\frac{\bar g_D\mu _q}{2}\right)^2\left[N^{-1}(T)+\bar f_{ l}^a+ \bar f_{ t}^a\right]^{-1} \label{eq:5.31.5}
\end{equation}
where $\bar f_{ l}^a$ and $\bar f_{t}^a$ denote the longitudinal and transverse parts of $\bar f^a$ respectively.
$\bar f^a_t$ has the logarithmic singularity, which comes from the OGE interaction between quarks
with collinear momenta. 
The longitudinal component $\bar f_{ l}^a$ has no IR singularity because
of the static screening and is almost temperature independent as
$f_{l;1}^s$ in Eq.~(\ref{NTl}).

The leading-order contribution at finite temperature comes from the
transverse component $\bar f_{ t}^a$; it has a logarithmic
singularity at $T=0$ due to the dynamic screening effect.
In this section, we shall see that 
the logarithmic divergences of
$N^{-1}(T)$ and $\bar f_{ t}^a$ at $T=0$ cancel out each other to
give a finite contribution to the magnetic susceptibility.
$\bar f_{ t}^a$ is given by
\begin{align}
 \bar f_{ t}^a=-2N_c N^{-1}(T)\int\frac{d^3k}{(2\pi)^3}\frac{\partial n(\epsilon_\bk)}{\partial \epsilon_{\bk}} \bar f^a_{{t};k,k_s} \label{eq:5.32}
\end{align}
with
\begin{align}
\bar f^a_{t:k,k_s} &=
 -\left.\int\frac{d\Omega_\bk}{4\pi}\int\frac{d\Omega_\bq}{4\pi}
 \frac{m^2}{E_s E_\bk}C_fN_c^{-1}g^2M^{iia} D_{ t}(k-q)\right|_{|\bq|=k_s}  \label{eq:5.35}
\end{align}
where $M^{iia}$ is the spin-dependent component of $M^{ii}$ in 
Eq.(7) (see Appendix A) and we have defined $E_s$ by $E_s=E_{|\bq|=k_s}$.
It is the dynamic screening part in the propagator $D_t$ that gives the
$\ln T$ dependence to $\bar f^a_{t}$. 
Therefore, we can put $|\bk|=k_s$ in the other parts of the integrand in Eq. (\ref{eq:5.35}).
Then,
\begin{equation}
M^{iia}=\frac{k_s^2}{4m^2}\left[1+\cos \theta_{\widehat{\bk\bq}}+\cos \theta_\bk \cos  \theta_\bq-\cos \theta_\bk^2 -\cos \theta_\bq^2\right].
\end{equation}
The real part of the transverse propagator is
\begin{align}
{\rm Re} D_{t}(k-q)\Big
 |_{|\bq|=k_s}&=\frac{(k-q)^2}{\left\{(k-q)^2\right\}^2+\left(\frac{1}{4}\sum_f \pi u_{F,f} 
m_{D,f}^2\right)^2\frac{(E_\bk-E_s)^2}{(\bk-\bq)^2}}\Bigg|_{|\bq|=k_s} \notag\\
&\simeq -\frac{1}{2k_s^2}\frac{(1-\cos \theta_{\widehat{\bk\bq}})^2}{(1-\cos \theta_{\widehat{\bk\bq}})^3+ c^3(E_\bk-E_s)^2}
\end{align} 
with  $c^3 \equiv
\frac{1}{8k_s^6}\left(\sum_f \frac{\pi m_{D,f}^2u_{F,f}}{4}\right)^2$, while the imaginary part is given by
\begin{align}
{\rm Im }D_{ t}(k-q)\Big|_{|\bq|=k_s} =
 \frac{2\sqrt{2}k_s^3c^{3/2}\frac{(E_\bk-E_s)}{|\bk-\bq|}}{\left\{(k-q)^2\right\}^2
+8k_s^6c^3\frac{(k_0-q_0)^2}{(\bk-\bq)^2}}\Bigg|_{|\bq|=k_s}.
\end{align}
It is easily shown that the imaginary part gives $\bar f^a$ only higher-order terms with respect to temperature and thus we neglect it here. 
Performing the angular integrals in Eq. (\ref{eq:5.35}),
\begin{multline}
\bar f^a_{t;k,k_s} 
=\frac{C_fN_c^{-1}g^2}{48E_s^2}\Big[ -10+2\ln \Big|\frac{c^3(E_\bk-E_s)^2+8}{c^3(E_\bk-E_s)^2} \Big| \\
+\frac{5c(E_\bk-E_s)^{2/3}}{6} \ln \Big|\frac{( c(E_\bk-E_s)^{2/3}+2)^3}{c^3(E_\bk-E_s)^2+8} \Big|  \\
+\frac{5c(E_\bk-E_s)^{2/3}}{6} \tan^{-1} \left(\frac{4-c(E_\bk-E_s)^{2/3}}{\sqrt{3} c(E_\bk-E_s)^{2/3}}\right)\\ 
+\frac{5\pi}{6\sqrt{3}} c(E_\bk-E_s)^{2/3}\Big]
\end{multline}
For small $|E_\bk-E_s|\sim O(T)$, it reads
\begin{align}
\bar f_{t;k,k_s}^a & \sim -\frac{C_fN_c^{-1}g^2}{8E_s^2}\ln
 \left[\frac{ c(E_\bk-E_s)^{2/3}}{2}
\right] \label{eq:fkTakinji}
\end{align}
The integral 
in Eq. (\ref{eq:5.32}) can be performed as in the case of $N(T)$, Eq. (20). 
Changing the variable of integration from $k$ to $\omega$
, we find
\begin{align}
\bar f_{t}^a \simeq \frac{N_c}{\pi^2 T} N^{-1}(T)\int_{\E_0}^\infty
 d\omega k(\omega)E_{k(\omega)}
\left(1-\frac{\partial {\rm Re}\Sigma_+(\omega)}{\partial \omega}\right)
 \bar f^a_{t;k(\omega),k_s} 
\frac{e^{\beta(\omega-  \mu)}}{\left(e^{\beta(\omega-  \mu)}+1\right)^2} . \label{eq:fat}
\end{align}
Note that $\d {\rm Re}\Sigma_+(\omega)/\d\omega$ in the integrand does
not contribute as long as we calculate to $O(g^2)$ as 
$\bar f^a_{k(\omega),t}$ itself is $O(g^2)$.  
Integrating Eq. (\ref{eq:fat}) as in Appendix C and using Eq. (\ref{eq:NTinv1}), we find a leading-order
contribution at $T\neq 0$, 
\begin{align}
\bar f_{t}^a &\sim N^{-1}(T)\frac{C_fg^2}{12\pi^2k_sE_s}\left[
1 +\frac{\pi^2}{6}\frac{(2k_s^2-m^2)}{k_s^4}T^2 \right]\ln T^{-1} +O(g^2T^2) \notag \\ 
&\sim \frac{C_fg^2}{12N_cE_s\mu}\ln T^{-1}. \label{eq:fas}
\end{align}

Compare Eq. (\ref{eq:fas}) with Eq. (\ref{eq:NTinv1}). Since
$E_s=E_F+O(T^2)$ and $k_s=k_F+O(T^2)$ as we shall see, the $\ln T$ terms cancel each other in the magnetic
susceptibility (\ref{eq:5.31.5}).
In Fig.1,  we plot $\pi^2/(N_ck_s\mu)-N^{-1}_t(T)$, $\bar f^a_t$, and $N^{-1}_t(T)+\bar f^a_t$ at $k_F=1.0$ fm$^{-1}$ near $T=0$. 
As is shown, both $N_t^{-1}(T)$ and $\bar f^a_t$ are singular at $T=0$, 
but a cancellation occurs so that the sum of the two
becomes finite even at $T=0$.    
Thus we can take the limit $T \rightarrow 0$ in the magnetic susceptibility. 

It is also worthwhile to note that $N_t(T)+\bar f^a_t$  seems to be almost temperature-independent in the unit of $\pi^2/(N_ck_s\mu)=1$ in Fig. 1. 
However, the chemical potential has the implicit temperature dependence.
We shall discuss this $T$-dependence in the following.

\begin{figure*}[h]
\begin{center}
\psfrag{N}{\hspace{-4em}$\frac{\pi^2}{(N_ck_s\mu)}-N^{-1}_t(T)$}
\psfrag{f}{\hspace{0em}{$\bar f^a_{t}$}}
\psfrag{S}{\hspace{-3em}$N^{-1}_t(T)+\bar f^a_{t}$}
\includegraphics[width=0.6\textwidth]{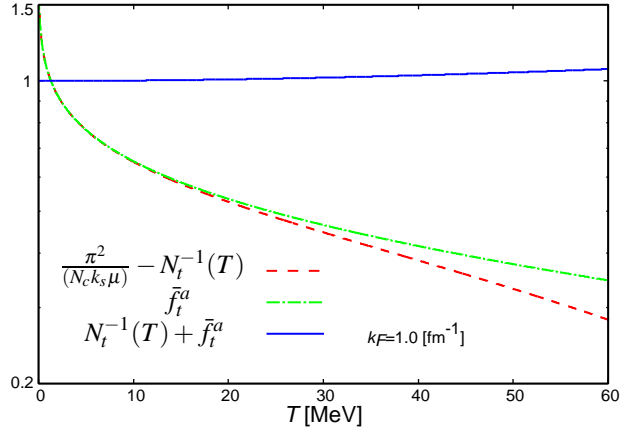}
\caption{Cancellation of the logarithmic terms. $\pi^2/(N_ck_s\mu)-N^{-1}_t(T)$, $\bar f^a_t$, and $N^{-1}_t(T)+\bar f^a_{t}$ are
 plotted for $\Lambda=M_D$ and $k_F=1.0$ fm$^{-1}$ in the unit of $\pi^2/(N_ck_s\mu)=1$.  $\bar f^a_t$ and
 $N^{-1}_t(T)$ are singular at $T=0$, but they cancel each other and the
 magnetic susceptibility remains finite even at $T=0$.}
\end{center}
\end{figure*}

\subsection{Magnetic susceptibility at T=0}

Before presenting the full expression of the temperature dependence, it
would be interesting to see how to recover the result at $T=0$
\cite{tat08,tat082}. 
We have seen that
the logarithmic terms in $N_t(T)^{-1}$ and $\bar f^a_t$ cancel each other
to leave the regular terms in the magnetic susceptibility at $T=0$. 
One can easily check that there arises no more singular terms from the
longitudinal contribution $\bar f^a_l$ at $T=0$, since the longitudinal
gluons are screened by the Debye mass $M_D$.  The Landau-Migdal
parameter $\bar f^a_l$ is given by 
\begin{align}
 \bar f_{ l}^a=-2N_c N^{-1}(T)\int\frac{d^3k}{(2\pi)^3}\frac{\partial
 n(\epsilon_\bk)}
{\partial \epsilon_{\bk}} \bar f^a_{l;k,k_s} \label{eq:5.46}
\end{align}
with
\begin{align}
\bar f^a_{l;k,k_s} &=
 -\left.\int\frac{d\Omega_\bk}{4\pi}\int\frac{d\Omega_\bq}{4\pi}
 \frac{m^2}{E_s E_\bk}C_fN_c^{-1}g^2M^{00a} D_{ l}(k-q)\right|_{|\bq|=k_s}  \label{eq:5.47}
\end{align}
in the same way as $\bar f^a_t$, where $M^{00a}$ is the spin-dependent component of $M^{00}$ in 
Eq.(7) (see Appendix A). Now the gluon propagator $D_l(k-q)$ has no infrared
singularity, so that 
\begin{align}
\bar f_{ l}^a&\simeq  \bar f^a_{l;k_s,k_s}\nonumber\\
&=
 -\left.\int\frac{d\Omega_\bk}{4\pi}\int\frac{d\Omega_\bq}{4\pi}
 \frac{m^2}{E_F^2}C_fN_c^{-1}g^2M^{00a} D_{
 l}(k-q)\right|_{|\bk|=|\bq|=k_F}+O(g^2 T^2).
\end{align}
Consequently $\bar f^a_l$ has little temperature dependence, and can be written
as 
\begin{align}
\bar
f^a_l&=-\frac{C_fg^2N_c^{-1}}{8E_F^2k_F^2}\left[2E_F^2I_0(\kappa)
+\frac{2}{3}\kappa(E_F-m)^2I_0(\kappa)\right.\nonumber\\
&+\left.\kappa \left\{k_F^2I_0(\kappa)+\frac{2}{3}\left((1+\kappa)I_0(\kappa)-1\right)(E_F-m)^2\right\}-k_F^2-\frac{2}{3}(E_F-m)^2\right]
\label{deno2}
\end{align}
with $\kappa=\sum_f m^2_{D,f}/2k_F^2$ at $T=0$ \cite{tat08,tat082}. 
Thus there is no singularity at $T=0$ in
the magnetic susceptibility (\ref{eq:5.31.5}). 

The magnetic susceptibility at $T=0$ is already given in
ref. \cite{tat08,tat082},
\begin{align}
&\hspace{1em}\left(\chi_M(T=0)/\chi_{\rm
Pauli}\right)^{-1} \notag\\
&=\frac{N_ck_F\mu_0}{\pi^2}\left[N^{-1}(0)+\bar
f^a_l+\bar f^a_t\right]\nonumber\\
&=
1-\frac{C_f g^2\mu_0 }{12 \pi^2 E_F^2k_F}\left[m(2E_F+m)-\frac{1}{2}\left( E_F^2+4E_Fm-2m^2\right)\kappa\ln \frac{2}{\kappa} \right], \label{eq:chi_pol}
\end{align}
where $\chi_{\rm Pauli}$ is the Pauli paramagnetism, $\chi_{\rm
Pauli}\equiv {\bar g}_D^2\mu_q^2N_ck_F\mu/4\pi^2$, and $\mu_0=E_F+O(g^2)$ is the
chemical potential at $T=0$. 
Using Eqs.~(\ref{eq:NTinv1}), (\ref{eq:fas}), (\ref{deno2}) for the
magnetic susceptibility (\ref{eq:5.31.5}) and comparing it with
Eq.~(\ref{eq:chi_pol}) at $T=0$, we find that they are different from each
other by the temperature-independent term $C_fg^2k_F/24\pi^2E_F$, which
is attributed to the contribution by the transverse gluons.



 Note that  the screening effect 
for the longitudinal gluons gives rise to the contribution of $g^4
\ln(1/g^2)$ \cite{her66,bru57}.
Obviously this expression (\ref{eq:chi_pol}) is reduced to the simple
OGE case without screening in the limit $\kappa \rightarrow 0$; one can see
that the interaction among massless quarks gives a null contribution
for the magnetic transition. 

\begin{figure*}[h]
\begin{center}
\includegraphics[width=0.6\textwidth]{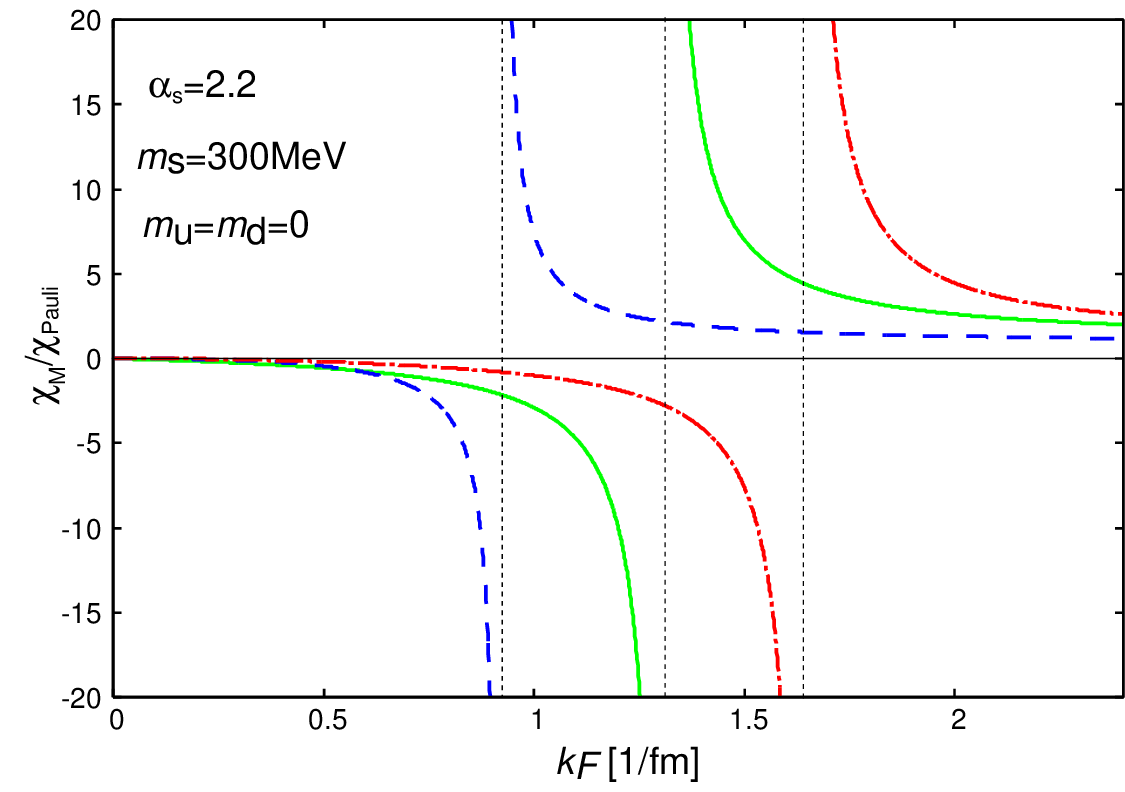}
\caption{Magnetic susceptibility at $T=0$. The solid curve shows the
 result using simple OGE, 
while the dashed and dash-dotted ones show the screening effects with $N_f=1$(only s quarks) and with $N_f=3$(u, d, and s quarks) respectively.}
\end{center}
\end{figure*}

In Fig. 2, we plot the magnetic susceptibility at $T=0$\cite{tat08,tat082}.
We 
take the QCD coupling constant as
$\alpha_s\equiv g^2/4\pi=2.2$ and the strange quark mass $m_s=300$MeV, inferred from the MIT bag model\cite{mit}. 
We consider here the MIT bag model as an effective model in our energy
scale, which succeeds in
reproducing the low-lying hadron spectra. The coupling constant looks
rather large, but this value is required for the color magnetic
interaction to explain the mass splitting
of hadrons with different spins; e.g. for nucleon and $\Delta$
isobar. We think this feature is relevant in our study, because 
the coupling constant is closely related to the strength
of the spin-spin interaction between quarks in this model. 
Moreover, the quark density in the MIT bag model is moderate,  
$\sim 0.25$fm$^{-3}$, which is the similar one we are here interested in.
Note that the perturbation method should be still meaningful even for this
rather large coupling, since the renormalization-group analysis has
shown that the relevant expansion parameter is not the gauge coupling
constant $g^2$ but the product of $g^2$ with the Fermi velocity $v_F$, 
which {\it always} goes to zero as one approaches to the Fermi surface \cite{sch}.  

One can see that the magnetic susceptibility for the simple OGE without screening
diverges around $k_F=1.3$ fm $^{-1}$.
This is consistent with the previous result for the energy calculation. \cite{tat08,tat082}.
One may expect that the screening effect weakens the Fock exchange interaction so that 
the critical density get lower once we take into account the screening effect. However, this is not
necessarily the case in QCD. The screening effect behaves in different ways depending on the number of flavors.
Compare the results for the $N_f=3$ with the one for $N_f=1$. In the case of $N_f= 1, \kappa \leq 2$ the screening effect 
works against the magnetic phase transition as in QED. 
However, for $N_f=3$, $\kappa>2$ so that the critical density is increased.
Consequently the screening effect does not 
necessarily work against the magnetic instability, which is a different
aspect from electron gas \cite{tat082}.
 
\subsection{Temperature dependence of the chemical potential}

The chemical potential $\mu$ in Eq. (\ref{eq:NTinv1}) implicitly includes the
temperature dependence. To extract the proper temperature dependence in
$\chi_M$ we must carefully take into account the temperature dependence
of $\mu$
; here we use the thermodynamic relation $\mu=-(\partial
F/\partial n)|_T$ with the free energy $F=E-Ts$. 

In the following we only consider the temperature variation $\delta T$,
so that the contribution of 
the longitudinal gluons can be well discarded due to the Debye
screening. Accordingly it is sufficient 
to take into account the contribution from the transverse gluons in Eq.~(\ref{eq:Sigma}). We shall see that $\mu$ 
includes $T^2\ln T$ term due to the dynamic screening effect for the transverse gluons, besides the usual $T^2$ term.   

Using the fact that 
 the temperature
variation of the free energy equals $-s\delta T$, 
we calculate the
quasi-particle entropy per unit volume $s$,
\begin{equation}
s=-N_c\sum_{\zeta}\int\frac{d^3k}{(2\pi)^3}\left[n(\bk, \zeta )\ln
					    n(\bk, \zeta )
+(1-n(\bk, \zeta ))\ln (1-n(\bk, \zeta ))\right]
\end{equation}
for each flavor, and then integrate it with respect to temperature to get $F$. 
The temperature variation of $s$ is given as 
\begin{equation}
\delta
 s=-\frac{1}{T}N_c\sum_\zeta\int\frac{d^3k}{(2\pi)^3}(\epsilon_{\bk,\zeta}-\mu)\delta n(\bk, \zeta )
\end{equation}
through the variation of the distribution function,
\begin{equation}
\delta n(\bk, \zeta )=\frac{\partial n(\bk, \zeta
 )}{\partial\epsilon_{\bk,\zeta}}
\left[-\frac{\epsilon_{\bk,\zeta}-\mu}{T}\delta T+\delta\epsilon_{\bk,\zeta}-\delta\mu\right].
\end{equation}
Since the term involving $\delta\epsilon_{\bk,\zeta}-\delta\mu$ gives at
least 
$T^3\ln T$ in the entropy at low
temperature \cite{bay04}, we shall see that the leading-order contribution is given
by the term with the explicit temperature variation $\delta T$,
\begin{equation}
\delta s=-N_c\sum_\zeta\int\frac{d^3k}{(2\pi)^3}\frac{\partial n(\bk,
 \zeta )}{\partial\epsilon_{\bk,\zeta}}
(\epsilon_{\bk,\zeta}-\mu)^2\frac{\delta T}{T}.
\end{equation}
Using Eq.~(\ref{jac}), $\delta s$ is recast into the following form,
\begin{equation}
\delta s=\frac{1}{T^3}\frac{N_c}{\pi^2}\int d\omega
 k(\omega)E_{k(\omega)}\left(1-\frac{\partial {\rm
			Re}\Sigma_+(\omega)}{\partial\omega}\right)
\frac{e^{\beta(\omega-\mu)}}{(e^{\beta(\omega-\mu)}+1)^2}(\omega-\mu)^2\delta T.
\end{equation}
Changing the variable $\omega$ by $y$ s.t. $\omega-\mu=yT$ and using Eq.~(\ref{eq:Sigma}), we have
\begin{eqnarray}
\delta s
=\frac{N_ck_FE_F}{3}\left(1-\frac{C_fg^2u_F}{12\pi^2}\ln\left(\frac{T}{\Lambda}\right)+O(g^2)\right)\delta T.
\end{eqnarray}
as in Appendix C.
Thus the entropy is
\begin{equation}
s=\frac{N_ck_FE_FT}{3}\left(1-\frac{C_fg^2u_F}{12\pi^2}\ln\left(\frac{T}{\Lambda}\right)\right)+O(g^2T). \label{eq:entropy}
\end{equation}
Note that this result is the same as the one given in ref.\cite{ipp}, and
the non-Fermi-liquid behavior of the specific heat $c_V=T(\partial
s/\partial T)_V=s+O(g^2 T)$ is also reproduced. The free energy $F$ is then given by the integral of $s$ with respect to temperature,
\begin{equation}
F=E_0+\frac{N_cE_Fk_F}{6}T^2+\frac{N_cC_fg^2k_F^2}{72\pi^2}T^2\ln\left(\frac{\Lambda}{T}\right)+O(g^2T^2),
\end{equation}
where $E_0$ is the ground state energy at $T=0$.
Through the thermodynamic relation $\mu=-(\partial F/\partial n)$ we finally obtain
\begin{equation}
\mu(T)=\mu_0-\frac{\pi^2}{6}\frac{2k_F^2+m^2}{k_F^2E_F}T^2\left(1+\frac{C_fg^2u_F}{12\pi^2}
\ln\left(\frac{\Lambda}{T}\right)\right)+O(g^2T^2).
\label{mut}
\end{equation}
Note that this relation is also given by considering the total number density at finite temperature \cite{ipp}.

\subsection{Temperature dependence of the magnetic susceptibility}
One can easily find that temperature-dependent terms of $E_s$ is the same as those of $\mu(T)$ within the order we are interested in,
noticing that $E_s=\mu-{\rm Re}\Sigma_+(\mu)$ and ${\rm Re} \Sigma_+(\mu)={\rm Re} \Sigma_+(\mu_0)+O(g^2T^2)$. 
Using $k_s=\sqrt{E_s^2-m^2}$, we find the temperature dependence of $k_s$,
\beq
k_s=k_F-\frac{\pi^2}{6}\frac{2k_F^2+m^2}{k_F^3}T^2\left(1+\frac{C_fg^2u_F}{12\pi^2}\ln\left(\frac{\Lambda}{T}\right)\right)+O(g^2T^2).\label{eq:ks}
\eeq

Now we can rewrite Eqs. (\ref{eq:NTinv1}) and (\ref{eq:fas}) 
in terms of $k_F$ and $E_F$;
\begin{align}
N^{-1}(T)&=\frac{\pi^2}{N_ck_F\mu_0}\Big[1+\frac{\pi^2}{6k_F^4}\left(2E_F^2-m^2+\frac{m^4}{E_F^2}\right)T^2 \notag \\
&-\frac{C_fg^2u_F}{72}\frac{2k_F^2-m^2}{k_F^4}T^2\ln\left(\frac{\Lambda}{T}\right)-\frac{C_fg^2u_F}{12\pi^2}\ln 
\left(\frac{\Lambda}{T}\right)\Big]-N_l(T)+O(g^2T^2) \label{eq:NTinv2}.
\end{align} 
and 
\beq
\bar f_{\rm t}^a 
=\frac{C_fg^2}{12N_cE_F\mu_0}\left[1+\frac{\pi^2}{3k_F^2E_F^2}\left(2k_F^2+m^2\right)T^2\right]\ln T^{-1}+O(g^2T^2). \label{eq:5.41}
\eeq

Taking into account this temperature dependence of the chemical
potential and using Eqs.~(\ref{eq:NTinv2}) and (\ref{eq:5.41}), we find the temperature dependence
of $\delta\chi_M^{-1}$,
\begin{align}
\delta\chi_M^{-1}&=\chi_{\rm Pauli}^{-1}\Big[ \frac{\pi^2}{6k_F^4}\left(2E_F^2-m^2+\frac{m^4}{E_F^2}\right)T^2 \notag\\
&\hspace{5em}+\frac{C_fg^2u_F}{72k_F^4E_F^2} \left(2k_F^4+k_F^2m^2+m^4\right)T^2\ln\left(\frac{\Lambda}{T}\right)\Big]+O(g^2T^2).
\label{delT}
\end{align}
Finally we obtain the magnetic susceptibility at finite temperature by
adding $\delta\chi^{-1}_M$ to Eq.~(\ref{eq:chi_pol}),
\begin{eqnarray}
\left(\chi_M/\chi_{\rm Pauli}\right)^{-1} = 
1 -\frac{C_fg^2}{12\pi^2E_Fk_F}\Big[m(2E_F+m)%
-\frac{1}{2}(E_F^2+4E_Fm-2m^2)
\kappa\ln\frac{2}{\kappa} \Big] \nonumber\\
+\frac{\pi^2}{6k_F^4} \left(2E_F^2-m^2+\frac{m^4}{E_F^2} \right)T^2
+\frac{C_fg^2u_F}{72}\frac{(2k_F^4+k_F^2m^2+m^4)}{k_F^4E_F^2} T^2\ln\left(\frac{\Lambda}{T}\right) \notag\\
+O(g^2T^2). 
\label{chi_finT}
\end{eqnarray}
There appears $T^2 \ln T$ dependence in the
susceptibility 
at finite temperature besides the usual $T^2$ dependence \cite{tat082}.
This corresponds to the $T \ln T$ term in the specific heat \cite{boy01,hol,rei,gan,cha,ipp,ger05} 
and is a novel {\it non-Fermi-liquid effect} in the magnetic susceptibility.
At low temperature, $\ln (\Lambda/T)>0$ so that the $T^2 \ln T$ term
gives positive contribution to 
$\chi_M^{-1}$.
Therefore, both $T$-dependent terms in Eq. (\ref{chi_finT}) work against the magnetic instability.

In Fig.3, we show the temperature dependence of the magnetic susceptibility at $k_F=1.0$ fm$^{-1}$ using Eqs.(\ref{delT}) and (\ref{chi_finT}). 
The thin solid, dash-dotted, heavy solid, and dashed curves indicate  
$\delta \chi_M(T)/\chi_{\rm Pauli}$, $\chi_M(T=0)/\chi_{\rm Pauli}$, $\chi_M(T)/\chi_{\rm Pauli}$ and $(k_F \mu_0)/(k_s\mu)$ for $\Lambda=M_D$ respectively.
The temperature at which the heavy solid curve cross the zero corresponds to the critical(Curie) temperature.  
Temperature dependence of the magnetic susceptibility, {\it i.e.} $\delta \chi_M$ originates from the transverse gluons.
$(k_F \mu_0)/(k_s\mu)$ is rewritten as
\beq
\left(\frac{\bar g_D\mu_q}{2}\right)^{-2}\frac{\pi^2}{Nk_s\mu}\Big/\chi^{-1}_{\rm Pauli}
\eeq
and corresponds to the first term in Eq.(\ref{eq:NTinv1}).
Once we take into account the temperature dependence of the chemical
potential, this term depends on temperature and adds 
temperature dependence other than the explicit temperature dependence to the magnetic susceptibility.  
We also plot the magnetic susceptibility with the temperature dependence of the chemical potential disregarded. 
If we ignore the temperature dependence of the chemical potential, the magnetic susceptibility hardly depend on temperature.
The temperature dependence of the magnetic susceptibility almost comes from that of the chemical potential.

\begin{figure*}[h]
\begin{center}
\psfrag{A}{\small \hspace{-4em}$\delta \chi^{-1}_M(T)/\chi^{-1}_{\rm Pauli}$}
\psfrag{B}{\small \hspace{-3em}$\chi^{-1}_M(0)/\chi^{-1}_{\rm Pauli}$}
\psfrag{D}{\small \hspace{-1em}$ \chi^{-1}_M(T)/\chi^{-1}_{\rm Pauli}$}
\psfrag{E}{\small \hspace{-1.5em}$(k_F\mu_0)/(k_s\mu)$}
\includegraphics[width=0.6\textwidth]{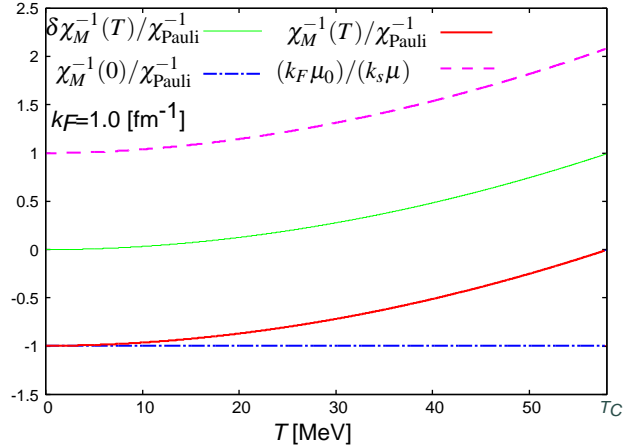}
\caption{Temperature dependence of the magnetic susceptibility at
 $k_F=1.0$ fm$^{-1}$. 
$\delta \chi^{-1}_M(T)/\chi^{-1}_{\rm Pauli}$,
 $\chi^{-1}_M(T=0)/\chi^{-1}_{\rm Pauli}$, 
$\chi^{-1}_M(T)/\chi^{-1}_{\rm Pauli}$, and $(k_F \mu_0)/(k_s\mu)$ are
 plotted for $\Lambda=M_D$. 
The dotted curve is the magnetic susceptibility in which the temperature dependence of the chemical potential is neglected.
The parameters are the same as those used in Fig. 1. 
$\chi^{-1}_M(T)$ changes its sign at $T=T_c$, which gives a critical (Curie) temperature.}
\end{center}
\end{figure*}

\section{Magnetic phase diagram of QCD on the density-temperature plane}

In this section, we show some numerical results for the magnetic
susceptibility at finite temperature.
We consider a symmetric quark matter with three flavors in equal populations, massless $u, d$ quarks and massive $s$ quarks.
We use the same values for $m$ and $\alpha_s$ as in Fig. 2.

\begin{figure}
\begin{center}
\includegraphics[width=0.6\textwidth]{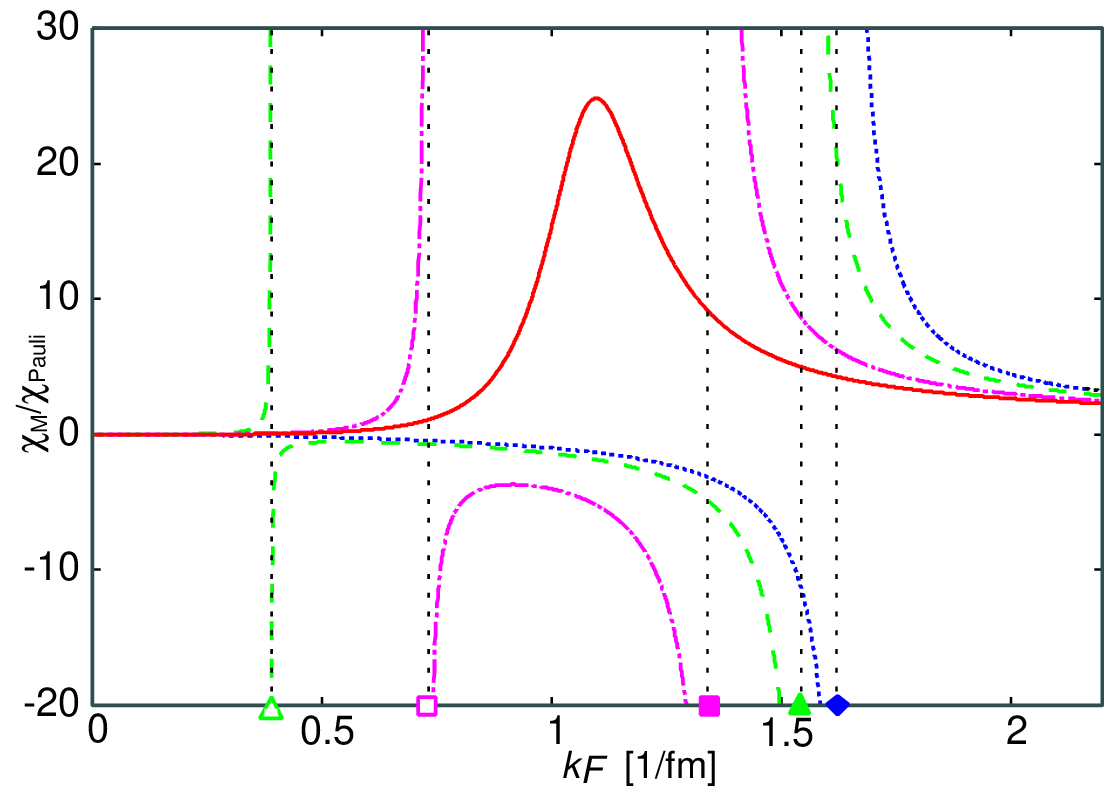}
\label{fig:chi}
\vspace{-1em}
\end{center} 
\begin{center}
(a)
\end{center}
\begin{center}
\includegraphics[width=0.6\textwidth]{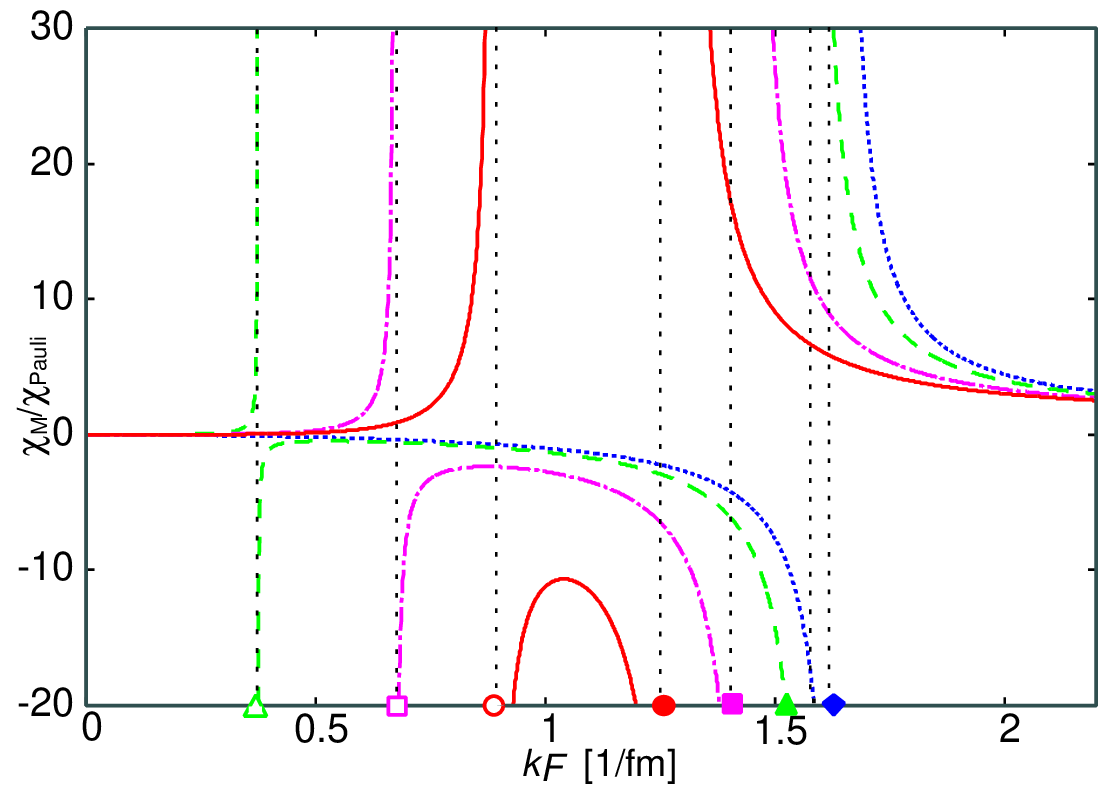}
 \label{fig:chi_wo_lnT}
 \vspace{-1em} 
\begin{center}
(b)
\end{center}
 \vspace{-1em} 
\caption{Magnetic susceptibility at finite temperature. The upper panel shows the result for the full expression, Eq.(\ref{chi_finT}),
 while the lower one shows the result without the $T^2 \ln T$ term in Eq.(\ref{chi_finT}). 
The dotted, dashed, dash-dotted, and solid curves show the results at $T$=0, 30, 50, and 60 MeV respectively.
The diamonds indicate the critical density, $k_F^c$ at $T=0$ and the
 open (filled) triangles, squares, and circles indicate $k_{F1}^c(k_{F2}^c)$ for $T=$30, 50, and 60MeV respectively.}
\end{center}
\end{figure}

In Fig.4(a), we plot the magnetic susceptibility given by Eq. (\ref{chi_finT}).
At $T$=0, the magnetic susceptibility is positive at higher densities
and the quark matter is in 
the paramagnetic phase there. At the critical density where the magnetic
susceptibility diverges ($k_F^c\sim 1.6$fm$^{-1}$), 
there occurs a magnetic phase transition from the paramagnetic phase to the ferromagnetic phase and the quark matter remains ferromagnetic below $k_F^c$.

At $T=$30 MeV, there appear two critical densities at which the magnetic
susceptibility  diverges. We denote these 
densities $k_{F1}^c$ and $k_{F2}^c$ ($k_{F1}^c<k_{F2}^c$). In this case,
$k_{F1}^c \simeq 0.4$fm$^{-1}$ and $k_{F2}^c \simeq 1.5$ fm$^{-1}$. 
At densities below $k_{F1}^c$ and above $k_{F2}^c$, the magnetic
susceptibility is positive, which corresponds to the paramagnetic phase, 
on the other hand, at densities between two critical densities, it becomes  negative corresponding to the ferromagnetic phase.

At $T=$50 MeV, there are still two critical densities ($k_{F1}^c\simeq
0.7$fm$^{-1}$ and $k_{F2}^c \simeq 1.3$fm$^{-1}$), 
but the range between these two densities becomes narrower than at $T=$30 MeV.

At $T=$60 MeV, there is no longer divergence in the magnetic susceptibility and the quark matter is paramagnetic at any density.   

To figure out the non-Fermi-liquid effect, in Fig.4(b), we depict the
magnetic 
susceptibility without the $T^2 \ln T$ term in Eq.(\ref{chi_finT}) for comparison: only $T^2$- dependence on temperature is taken into account.
At $T=0$, there is no difference from the curve shown in Fig.4(a) and there is little difference between Fig.4(a) and (b) even at $T=$30 and 50 MeV.
However, at $T=$60MeV, there is a region where the spontaneous
magnetization is still realized in Fig.4(b), 
whereas the quark matter is paramagnetic at any density in Fig. 4(a).   
From these considerations, the non-Fermi-liquid effect lowers the Curie
temperature to some extent and affects 
the magnetic property of quark matter in contrast with that of the anomalous specific heat ($\propto T\ln T$) in a non-relativistic electron gas~\cite{hol,rei,gan,cha}.   


We show a magnetic phase diagram of QCD on the density-temperature plane in Fig.5.
The four curves corresponds to the critical curves given by
Eq.(\ref{chi_finT}) 
under four different assumptions: below the curves the quark matter is
in the ferromagnetic phase, 
while it is in the paramagnetic phase above the critical curves. The magnetic transition occurs on the critical curves.

For the solid curve, we have used the full expression
Eq.(\ref{chi_finT}), on the other hand, 
for the dashed, dash-dotted, and dotted curves, we have ignored the
dynamic screening({\it i.e.} the $T^2 \ln T$ term), 
static screening({\it i.e.} the $\kappa \ln \kappa$ term), and both of the two screenings in Eq. (\ref{chi_finT}) respectively.
 
Compare the result with the full expression (\ref{chi_finT}) with the
one without 
the non-Fermi-liquid effect {\it i.e.} $T^2 \ln T$ dependence. In the
case without the $T^2 \ln T$ term, 
the ferromagnetic phase can be sustained till over $T=60$ MeV, while it can be at most $T=60$MeV including $T^2 \ln T$ dependence.
It turns out that the dynamic screening works against the magnetic
instability and can reduce the ferromagnetic region in the phase diagram up to a point,
but this effect is not so large.

The dash-dotted curve is the result without the static screening or
$\kappa \ln \kappa$ term in Eq.(\ref{chi_finT}). The static screening
effect works in favor of the magnetic instability to enlarge the
ferromagnetic region. As discussed in ~\cite{tat082}, it depends on the
number of flavors whether the static screening works for the 
ferromagnetism or not, which is peculiar to QCD.

\begin{figure*}[h]
\begin{center}
\includegraphics[width=0.8\textwidth]{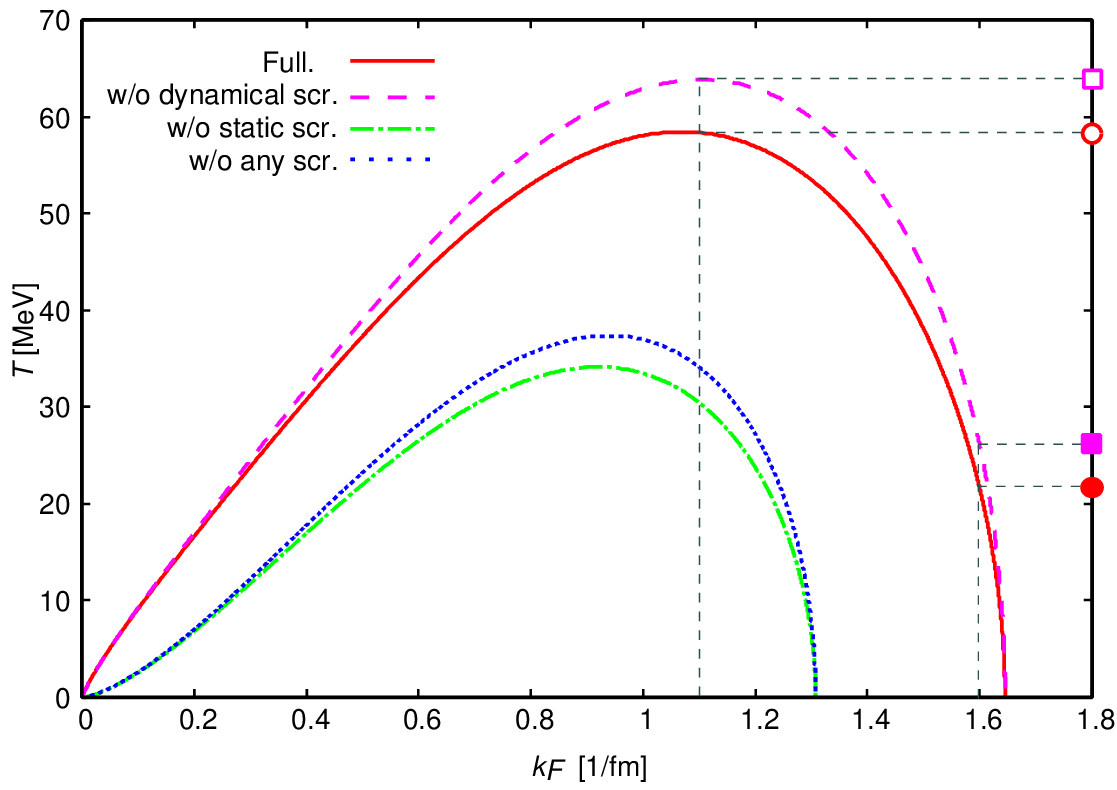}
\caption{Magnetic phase diagram in the density-temperature plane. The
 solid, dashed, dash-dotted, dotted curves show the results for the full
 expression Eq. (\ref{chi_finT}), the one without the $T^2 \ln T$ term,
 without the $\kappa \ln \kappa$ term, and without the $T^2 \ln T$ and
 $\kappa \ln \kappa$ terms in Eq. (\ref{chi_finT}). The open (filled)
 circle indicates 
the Curie temperature at $k_F=1.1(1.6)$ fm$^{-1}$ while the squares show those when we disregard the $T^2 \ln T$ dependence.}
\label{Fig:diagram}
\end{center}
\end{figure*}

\begin{figure}[tbp]
\vspace{-5em}
\begin{center}
\subfigure[]{
\psfrag{A}{\small $\delta \chi_{T^2}^{-1}$}
\psfrag{B}{\small \hspace{-1em}$\delta \chi_{T^2\ln T}^{-1}$}
\psfrag{D}{\small $-\chi_M^{-1}(0)$}
\psfrag{F}{\hspace{-1em}$\delta \chi^{-1}_{M}(T)$}
\psfrag{G}{}
\includegraphics[width=0.48\textwidth]{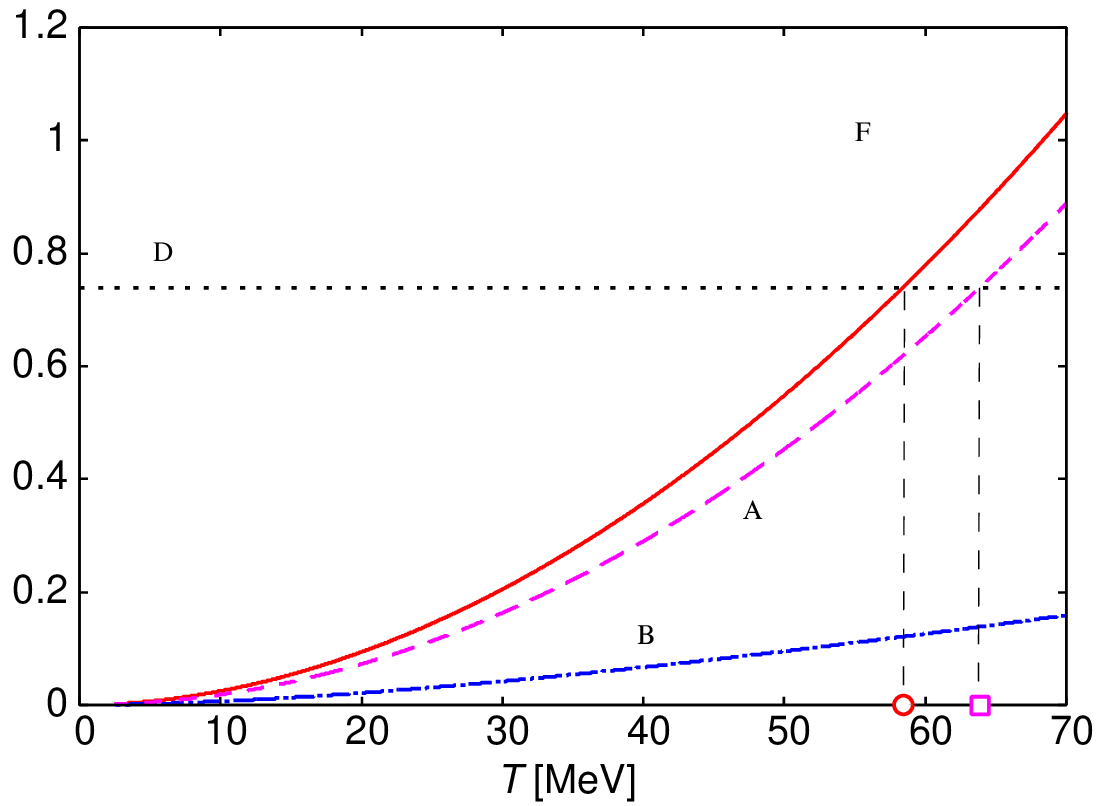}
 \label{}}
\subfigure[]{
\psfrag{A}{\small $\delta \chi_{T^2}^{-1}$}
\psfrag{B}{\small\hspace{-2em}$\delta \chi_{T^2\ln T}^{-1}$}
\psfrag{D}{\small  $-\chi_M^{-1}(0)$}
\psfrag{F}{\hspace{-1em}$\delta \chi^{-1}_{M}(T)$}
\psfrag{G}{}
\includegraphics[width=0.48\textwidth]{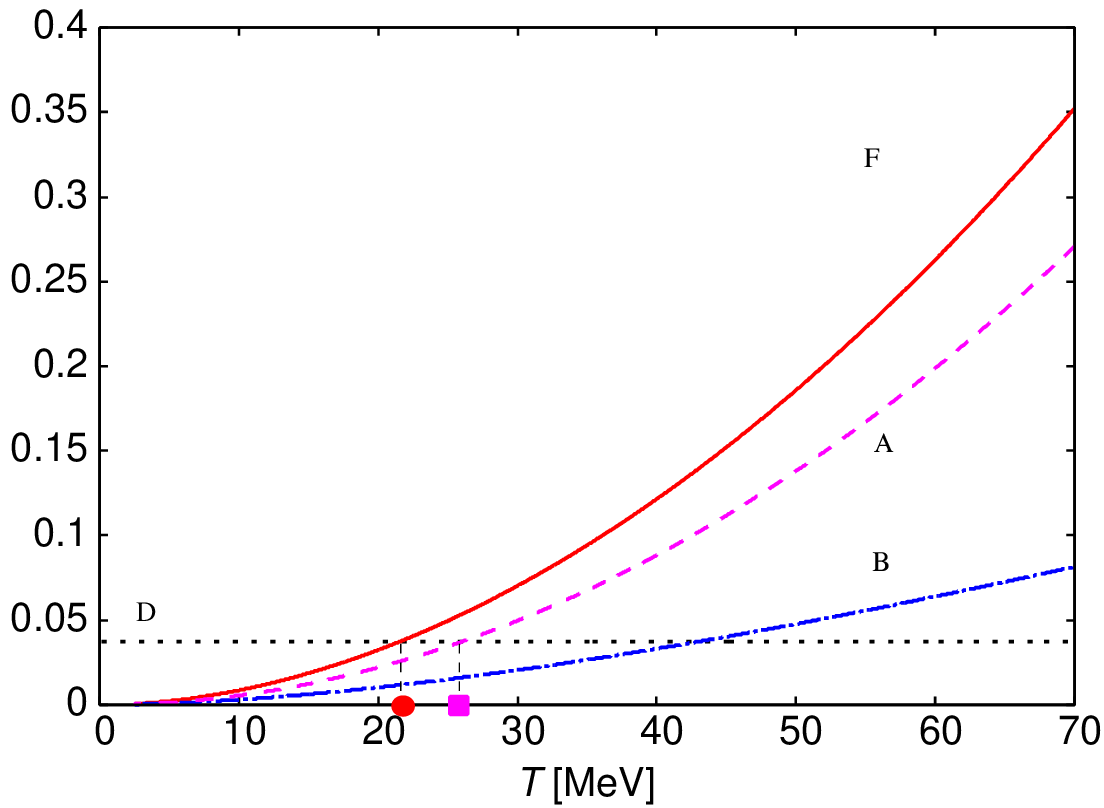}
\label{}}
\end{center}
\vspace*{-8mm}
\caption{Contributions to the magnetic susceptibility. The left panel shows the result for $k_F=1.1$ fm$^{-1}$
 and the right one is for $k_F=1.6$ fm$^{-1}$ in the unit of $\chi_{\rm Pauli}=1$. 
The solid, dashed, dash-dotted, and dotted curves show $\delta \chi^{-1}_{M}(T)$, $\delta \chi_{T^2}^{-1}$, $\delta \chi_{T^2\ln T}^{-1}$,
and $-\chi_M^{-1}(0)$ respectively.
The circles(squares) indicate the critical(Curie) temperatures including(ignoring) the $T^2 \ln T$ dependence. 
$\delta \chi^{-1}_{T^2 (T^2\ln T)}$ means the $T^2$ ($T^2\ln T$) term in the magnetic susceptibility, Eq. (\ref{chi_finT}).} 
\label{contribution}
\end{figure}
 
The maximum Curie temperature $T_c^{\rm max}$ is around $60$MeV, which is achieved at $k_F\simeq
1.1$fm$^{-1}$. Note that this is still low temperature, since $T_c^{\rm
max}/k_F\ll 1$. Thus our low-temperature expansion is legitimate over all
points on the critical curve. One of interesting phenomenological
implications may be related to thermal evolution of magnetars;
during the supernova expansions
temperature rises up to several tens MeV, which is so that ferromagnetic phase
transition may occur in the initial cooling stage to produce huge
magnetic field.

In Fig.\ref{contribution}, we plot the contributions to the magnetic susceptibility from the temperature independent, $T^2$ ,  and
$T^2\ln T$ terms for $k_F=1.1$ fm$^{-1}$ and $k_F=1.6$ fm$^{-1}$. 
We represent the $T^2$ and $T^2 \ln T$ term in Eq. (\ref{chi_finT}) by $\delta \chi_{T^{2}}^{-1}$ and $\delta \chi_{T^{2}\ln T}^{-1}$ respectively. 
Compare the contributions from the two temperature-dependent terms.
In either of the cases, the $T^2$ term is dominated over the $T^2\ln T$ term except at extremely low temperature.
At sufficiently low temperature, the $T^2 \ln T$ term should exceed the $T^2$ term, although we cannot see such a situation in Fig.6.  
Now, we estimate the reversal temperature, $T_{rev}$, at which $\delta \chi_{T^{2}}^{-1}$ starts to exceed $\delta \chi_{T^{2}\ln T}^{-1}$.
This temperature is easily obtained from Eq. (\ref{delT}), 
\begin{align}
T_{rev}&=\Lambda \exp \left[-\frac{12\pi^2}{C_fg^2u_F}\left(1+\frac{2k_F^2m^2+m^4}{2k_F^4+k_F^2m^2+m^4}\right) \right] \notag\\
       &=T_{rev,c}\exp \left[-\frac{12\pi^2}{C_fg^2u_F}\left(\frac{2k_F^2m^2+m^4}{2k_F^4+k_F^2m^2+m^4}\right) \right]
\end{align}
where $T_{rev,c}$ is the reversal temperature of the specific heat, at which  the $T$ term and the $T \ln T$ term in the specific heat
or the entropy, Eq.(\ref{eq:entropy}) are equal in magnitude.
$T_{rev}$ is quite low for the densities we are interested in.(For example, $T_{rev}=O(0.01)$MeV for $k_F=1.1$ fm$^{-1}$ and    $T_{rev}=O(1)$ MeV for $k_F=1.6$ fm$^{-1}$.)
As is shown in Fig. 6, the $T^2 \ln T$ dependence is too important to neglect, although the reversal temperature of the magnetic susceptibility is quite low.

\section{Summary and concluding remarks}

We have discussed some magnetic aspects of quark matter by evaluating
the magnetic susceptibility at finite temperature based on
QCD, where the screening effects for gluons are taken into account.
Different from the usual FLT, we cannot use the low temperature
expansion because the quasi-particle energy have an anomalous term on the
Fermi surface.
Carefully studying the temperature effects, we have found that the
magnetic 
properties of QCD exhibit interesting
features reflecting the gauge interaction: the dynamic screening of the 
transverse gluon field gives 
an anomalous $T^2 \ln T$ contribution 
to the magnetic susceptibility \cite{tat083}, while the contribution by
the longitudinal gluons is almost temperature independent due to the
Debye mass.
It may be interesting to recall that the dynamic screening has no effect but the static screening gives 
the term proportional to $M_D^2
\ln M_D^{-1}$ at $T=0$\cite{tat082} with the Debye mass $M_D$; 
$M_D$ works as an infrared (IR) cutoff to 
remove the infrared singularity in
the quasi-particle interaction exchanging the longitudinal gluons, while
the IR singularity still remains in the quasi-particle
interaction due to the transverse gluons. At
finite temperature, the energy transfer of order $T$ is allowed among the
quasi-particles in the vicinity of the Fermi surface, so that temperature 
itself plays a role of the IR cutoff through the dynamic
screening for the quasi-particle interaction.   
The logarithmic temperature dependence appears in the magnetic susceptibility  
as a novel non-Fermi-liquid effect, and its origin is the same as in the well-known 
$T\ln T$ dependence of the specific heat~\cite{boy01,hol,rei,gan,cha,ipp,ger05}.   
The dynamic screening also gives $\ln T$ terms for the spin-asymmetric
Landau-Migdal parameter as well as the density of states on the Fermi
surface as the leading-order contribution, 
but they exactly cancel each other in the formula of the magnetic susceptibility. Hence 
$T^2\ln T$ term becomes the leading-order contribution.
The anomalous $T^2 \ln T$ term works against the magnetic
instability, as well as the usual $T^2$ term. 

We demonstrated how the temperature effects work by drawing the magnetic phase diagram 
on the density-temperature plane. We also figured out the role of the non-Fermi-liquid effect 
in the phase diagram. 
We have seen that the $T^2 \ln T$ term should have a sizable effect on
the magnetic phase diagram in the temperature-density plane
\footnote{It should be worth noting that the anomalous $T^2\ln T$ term
in the magnetic susceptibility could be also observed in electron gas. }
.

We have seen some relations between the microscopic many-body
calculation and the FLT; the Landau-Migdal parameter $f_1^s$ or $F_1^s$ for the spin
symmetric interaction is closely
connected with the quark self-energy and thereby the renormalization
factor $z_+(k_F)$. It diverges at $T=0$ due to the transverse mode, but
another divergence coming from the spin dependent interaction cancel
it to give a finite result for the magnetic susceptibility. At finite
temperature all the Landau-Migdal parameters become finite. 
We have also derived the temperature dependence of the chemical
potential within FLT to get the same result as that given by the
microscopic calculation.

The magnetic susceptibility is a powerful tool to study not only the magnetic
properties of QCD but also the magnetic transition in quark matter. 
However, it only gives the information about the curvature at the origin of 
the effective potential (free energy) with respect to magnetization. 
We have discussed the magnetic phase diagram in QCD, 
assuming that the magnetic transition is of the second order or the
weakly first order. 
If the phase transition is of the first-order, we must explore the
global behavior of the effective potential. The critical density or the
critical (Curie) temperature then would be
larger than that in this paper. It would be an interesting possibility to be further investigated\cite{inu,pal}.

Here we have only considered the Fock exchange interaction in a
perturbation method 
to reveal the interesting features of the magnetic properties within
gauge theories. We have used a rather large coupling constant to
demonstrate the specific features of the magnetic phase transition and
the magnetic phase diagram, referring to
the bag model parameter, where the color magnetic interaction gives a
spin-spin interaction to split the mass degeneracy, e.g. for nucleon and
$\Delta$ isobar. 
To be more realistic, we must take into account the
non-perturbative effects, since some non-perturbative effects, like
instantons \cite{inu,nam}, are still relevant at low and intermediate
densities. Actually some part of the mass splitting should be attributed to 
such non-perturbative effects, which in turn affect the
magnetic properties of quark matter. Although the coupling constant
should be small in this case, the non-Fermi-
liquid behavior still survives as a qualitative effect.  

In high-density QCD color superconductivity (CSC) has been extensively
studied by many authors \cite{csc}. We have completely discarded its
possibility in this paper to concentrate in the magnetic properties.
It would be interesting to study the interplay of superconductivity and
magnetism in quark matter. A first attempt in this direction has been already
done \cite{nak03}, but more studies are needed to this end.   

Finally, it should be worth noting that magnetic properties of 
QCD or its magnetic
instability may be related to physics of compact stars, especially
magnetars \cite{mag} or primordial
magnetic field in early universe, where one may expect the QCD phase
transition \cite{uni}.  Our phase diagram may have a direct relevance to
the thermal evolution of magnetars; the evolution path can be plotted on the
temperature-density plane, and it may traverse the critical line in the
initial cooling stage just after the supernova explosion.

\section*{Acknowledgment}

This work was partially supported by the 
Grant-in-Aid for the Global COE Program 
``The Next Generation of Physics, Spun from Universality and Emergence''
from the Ministry of Education, Culture, Sports, Science and Technology
(MEXT) of Japan  
and the 
Grant-in-Aid for Scientific Research (C) (16540246, 20540267).

\section*{Appendix A}

The matrix elements for two quarks with $\zeta$ and $\zeta'$
\beqa
M^{00}&=&{\rm
tr}[\gamma^0\rho(k,\zeta)\gamma^0\rho(q,\zeta')],\nonumber\\
M^{ij}&=&{\rm tr}[\gamma^i\rho(k,\zeta)\gamma^j\rho(q,\zeta')],
\eeqa  
can be easily evaluated.
A straightforward manipulation, using Eq.~(3), gives 
\beqa
M^{00}&=&\frac{1}{4m^2}\left\{2k_0q_0-k\cdot q+m^2+(m^2-q\cdot k)(2a_0b_0-a\cdot
b)\right.\nonumber\\
&-&\left.\left[-2a_0q_0k\cdot b+2k_0q_0a\cdot b-2k_0b_0a\cdot
q+k\cdot ba\cdot q\right]\right\},
\label{elel}
\eeqa
and
\beqa
M^{ij}
&=&\frac{1}{4m^2}\left\{(1-a\cdot b)\widehat{k^iq^j}+(m^2-k\cdot
q)\widehat{a^ib^j}+a\cdot q\widehat{k^ib^j}+b\cdot
k\widehat{q^ia^j}\right.\nonumber\\
&&+\left.g^{ij}\left[(m^2-k\cdot q)(1-a\cdot b)-k\cdot bq\cdot a\right]\right\},\label{eq:Mii}
\label{elet}
\eeqa
with a symbol, $\widehat{k^iq^j}=k^iq^j+k^jq^i$, where the spin vector
$a_\mu$ is given by Eq.~(2) and similarly 
$b_\mu$ by ${\bf b}=\bzeta'+\bq(\bzeta'\cdot\bq)/m(E_q+m), b^0=\bq\cdot\bzeta'/m$.

\section*{Appendix B}
We present here some relations between the self-energy $\Sigma_+$ and the Landau parameter.
The Fermi-liquid interaction of OGE is calculated at $T=0$ as 
\beq
\left.f_{\bk\zeta,\bq\zeta'}\right|_{|\bk|=|\bq|=k_F}=-\frac{m^2N_c^{-1}C_fg^2}{E_F^2}
\lim_{\omega\rightarrow\mu}\left[-M^{00}{
D}_l(\omega-\mu, \bk-\bq)+M^{ii}{D}_t(\omega-\mu, \bk-\bq)\right],
\eeq
where the matrix elements $M^{\mu\nu}$ are given in Appendix A, and the
propagator renders  
\beq
{D}_l(\omega-\mu,\bk-\bq)\simeq -\frac{1}{2k_F^2(1-\cos\theta_{\widehat{\bk\bq}}+\kappa)},
\eeq
for the longitudinal gluons with $\kappa$ being the parameter in terms of
the Debye mass, $\kappa=M_D^2/2k_F^2=\sum_f m_{D,f}^2/2k_F^2$. The propagator for the
transverse gluons renders
\beq
\left.D_t(\omega-\mu,\bk-\bq)\right|_{|\bk|=|\bq|=k_F}\simeq-\frac{1}{2k_F^2\left[1-\cos\theta_{\widehat{\bk\bq}}
+ic^{3/2}\frac{\omega-\mu}{(1-\cos\theta_{\widehat{\bk\bq}})^{1/2}}\right]},
\eeq
with $c^3 \equiv
\frac{1}{8k_s^6}\left(\sum_f \frac{\pi m_{D,f}^2u_{F,f}}{4}\right)^2$
 in the soft-gluon limit, $|\omega-\mu|/\mu\ll 1$.

It is rather easy to see the contributions to the Landau-Migdal parameters from the longitudinal
gluons, since they have no IR singularity. For the spin-averaged interaction,
we have 
\beq
f^s_l=-N_c^{-1}C_fg^2\frac{m^2}{E_F^2}\lim_{\epsilon_\bk\rightarrow
\mu}M^{00s}
D_l(\epsilon_\bk-\mu, \bk-\bq)|_{|\bk|=|\bq|=k_F},
\eeq  
with 
\beq
M^{00s}=\frac{1}{4}\sum_{\zeta,\zeta'}M^{00}=\frac{1}{4m^2}\left(E_F^2+k_F^2\cos\theta_{\widehat{\bk\bq}}+m^2\right).
\eeq
Accordingly the Landau-Migdal parameter, $f_{l;1}^s$ reads 
\begin{eqnarray}
f_{l;1}^s&\equiv&\frac{3}{2}\int
 d(\cos\theta_{\widehat{\bk\bq}})\cos\theta_{\widehat{\bk\bq}}f^s_l\nonumber\\
&=&-\frac{3N_c^{-1}C_fg^2}{8E_F^2k_F^2}\left[\kappa k_F^2+2E_F^2\right]
\left[(1+\kappa)I_0(\kappa)-1\right].
\label{f1l}
\end{eqnarray}
Using the relation of the density of state near the Fermi surface and
the spin-averaged Landau-Migdal parameter \cite{bay76}, we find
$1/3f_{l;1}^s$ contributes to $N(0)$ at $T=0$. Since there is little
temperature dependence for the short-range interaction, we have  $
N_l(T)\simeq N_ck_FE_F/(3\pi^2)f_{l;1}^s$. 
Recalling Eq.~(\ref{eq:Sigma}),
we can make sure that the longitudinal contribution is implicitly included in
$\Sigma_+$ through $\Delta^{\rm reg}$.

For the transverse gluons, we must carefully calculate the Landau-Migdal
parameters, since they never receive the static screening but the
dynamic one due to the Landau damping.
Generally $D_t$ includes the imaginary part proportional to the squared 
Debye mass, but it only gives a higher-order contribution. 
The real part of the Fermi-liquid interaction now
reads
\begin{equation}
\left.f_{t;\bk\zeta,\bq\zeta'}\right|_{|\bk|=|\bq|=k_F}\sim -\frac{N_c^{-1}C_fg^2m^2}{E_F^2}\lim_{\omega\rightarrow
\mu}\left[M^{ii}{\rm Re}D_t(\omega-\mu,\bk-\bq)\right]_{|\bk|=|\bq|=k_F}.
\end{equation}
For the spin-dependent Landau parameter $\bar f^a_t$ we have already
seen in \S 3 that 
\begin{equation}
\bar f^a_t\sim \frac{C_fN_c^{-1}g^2}{12E_F^2}\lim_{\epsilon_\bk\rightarrow \mu}{\rm ln}|\epsilon_\bk-\mu|
\end{equation}
as a leading-order contribution.
For the spin averaged interaction,
\begin{equation}
f^s_t=-N_c^{-1}C_fg^2\frac{m^2}{E_F^2}\lim_{\epsilon_\bk\rightarrow \mu}M^{iis}{\rm
Re}D_t(\epsilon_\bk-\mu, \bk-\bq)|_{|\bk|=|\bq|=k_F},
\end{equation}
with
\begin{equation}
M^{iis}\equiv\frac{1}{4}\sum_{\zeta,\zeta'}M^{ii}=\frac{k_F^2}{2m^2}\left[1+\frac{1}{2}(1-\cos\theta_{\widehat{\bk\bq}})\right].
\end{equation}
One can see that the relevant Landau parameter $f_{t;1}^s$ also 
contains ${\rm ln}|\epsilon_\bk-\mu|$ due to the angular integral. 
$f_{t;1}^s$ is defined by 
\begin{eqnarray}
f_{t;1}^s\equiv\frac{3}{2}\int d(\cos\theta_{\widehat{\bk\bq}})\cos\theta_{\widehat{\bk\bq}}f^s_t.
\end{eqnarray}

Consider the integral,
\begin{align}
&\hspace{1em}\int_{-1}^1 d(\cos\theta_{\widehat{\bk\bq}})\cos\theta_{\widehat{\bk\bq}} {\rm Re}D_t M^{iis} \notag\\
&=-\frac{1}{8m^2}\int_0^2(1-x)(2+x)\frac{x^2}{x^3+{\tilde c}^3}dx\nonumber\\
&=-\frac{1}{8m^2}\left[\frac{2}{3}\ln\left|\frac{8+{\tilde c}^3}{{\tilde c}^3}\right|
-4+\frac{{\tilde c}}{6}\ln\left|\frac{(2+\tilde c)^3}{8+{\tilde c}^3}\right|\right. 
+\frac{{\tilde c}}{\sqrt{3}}\left(
\tan^{-1}\frac{4-{\tilde c}}{\sqrt{3}{\tilde c}}+\frac{\pi}{6}\right) \notag\\
&\hspace{12em}+\frac{{\tilde c}^2}{6}\ln\left|\frac{8+{\tilde c}^3}{(2+{\tilde c})^3}\right| 
+\left.\frac{{\tilde c}^2}{\sqrt{3}}\left({\rm
tan}^{-1}\frac{4-{\tilde c}}{\sqrt{3}{\tilde c}}+\frac{\pi}{6}\right)\right],
\end{align}
with $\tilde c=c(\epsilon_\bk-\mu)^{2/3}$. For a small $|\epsilon_\bk-\mu|$, it reads
\begin{equation}
= -\frac{1}{8m^2}\left[-2{\rm ln}\left(\frac{c(\epsilon_\bk-\mu)^{2/3}}{2}\right)-4
+O((\epsilon_\bk-\mu)^{4/3})\right].
\end{equation}

Thus the leading-order contribution to $f_{t;1}^s$ is eventually given by 
\begin{eqnarray}
f_{t;1}^s\equiv \lim_{\epsilon_\bk\rightarrow
 \mu}f_{t;1\bk}^s,
\label{f1s}
\end{eqnarray}
with $f_{t;1\bk}^s\sim -\frac{C_fg^2N_c^{-1}}{4E_F^2}{\rm ln}|\epsilon_\bk-\mu|$
or
\beq
F_{t;1}^s\equiv N(0)f_{t;1}^s\simeq -\frac{C_fg^2k_F}{4\pi^2E_F}\lim_{\epsilon_\bk\rightarrow
 \mu}
{\rm ln}|\epsilon_\bk-\mu|,
\label{F1s}
\eeq
up to $O(g^2)$. Comparing Eq.~(\ref{F1s}) with Eq.~(\ref{eq:Sigma}), we can see
\beq
F_{t;1}^s=\lim_{\epsilon_\bk\rightarrow \mu}
\left.\frac{\partial{\rm Re}\Sigma_+(\omega)}{\partial\omega}\right|_{\omega=\epsilon_\bk},
\eeq
in the leading order.

Consider the renormalization factor on the Fermi surface, which measures a
discontinuity of the momentum distribution of the bare particles
\cite{noz,lut,lut61}. We
find 
\begin{equation}
z_+(k_F)\simeq\left(1+\frac{1}{3}F_{t;1}^s\right)^{-1}=0. 
\label{ren}
\end{equation}

The Fermi velocity is given by 
\begin{equation}
v_F^{-1}\simeq\frac{\mu}{k_F}\left(1+\frac{1}{3}F_{t;1}^s\right)\rightarrow \infty.
\end{equation}
Moreover, one can see that 
\begin{equation}
v_F=\frac{k_F}{\mu}z_+(k_F).
\end{equation}
In the non-relativistic version, this relation is nothing but the
relation between the effective mass $m^*$ and the renormalization factor
$z_+(k_F)$, $m^*=z_+^{-1}(k_F)$.

At $T\neq 0$, the contribution of $f_{t;1}^s$ to $N(T)$ can be estimated by
evaluating 
\beq
-2N_c\int\frac{d^3k}{(2\pi)^3}\frac{\partial
n(\epsilon_\bk)}{\partial\epsilon_\bk}\frac{f_{t;1\bk}^s}{3}
=-\frac{C_fg^2u_F}{12\pi^2}\ln T,
\label{ntf}
\eeq
which is reduced to $1/3F^s_{t;1}$ at $T=0$.
Thus we can see that the $\ln T$ term in Eq.~(\ref{eq:NT1}) is just given by the
Landau-Migdal parameter $f_{t;1}^s$.

\section*{Appendix C}

Here we derive Eq.~(\ref{eq:NT1}). Substituting Eq.~(\ref{eq:Sigma}) into Eq.~(\ref{eq:NTomega}) and changing the
variable $\omega$ by the dimensionless one $y$ through $\omega-\mu=yT$,
we have
\beq
N(T)\simeq\frac{N_c}{\pi^2}\int_{-\infty}^{\infty} dy k(\mu+yT)E_{k(\mu+yT)}
\left(1-\left.\frac{{\rm
Re}\Sigma_+(\omega)}{\partial\omega}\right|_{\omega=\mu+yT}\right)
\frac{e^y}{(e^y+1)^2},
\eeq
where we put $(\epsilon_0-\mu)/T\rightarrow -\infty$ in the lower limit
of the integral.
Now we expand the each term in the integrand to find the leading-order
contribution at low temperature. We thus have 
\beq
1-\left.\frac{\partial{\rm
Re}\Sigma_+(\omega)}{\partial\omega}\right|_{\omega=\mu+yT}\simeq
1-\frac{C_fg^2u_F}{12\pi^2}
\ln\left(\frac{T}{\Lambda}\right),
\eeq
and
\beq
k(\mu+yT)E_{k(\mu+yT)}\simeq k_sE_s+\frac{E_F(2k_F^2-m^2)}{2k_F^3}y^2T^2\left(1-\frac{C_fg^2u_F}{6\pi^2}
\ln\left(\frac{T}{\Lambda}\right)\right),
\eeq
in the integrand, where the odd-power terms of $y$ is discarded
since they never contribute to the integral with respect to $y$.
Finally we have
\beqa
\!\!\!\!\!\!\!\!\!\!\!N(T)\simeq \frac{N_c}{\pi^2}\int_{-\infty}^\infty dy\!\!\!\!\!\!\!&&\!\!\!\!\!\!\left[k_sE_s\left(1-\frac{C_fg^2u_F}{12\pi^2}
\ln\left(\frac{T}{\Lambda}\right)\right)\right.\nonumber\\
&&\!\!\!+\!\!\!\left.\frac{E_F(2k_F^2-m^2)}{2k_F^3}y^2T^2\left(\!\!1-\frac{C_fg^2u_F}{4\pi^2}
\ln\left(\frac{T}{\Lambda}\right)\right)
\right]\frac{e^y}{(e^y+1)^2}.
\label{ant}
\eeqa
The integral in Eq.~(\ref{ant}) is easy to give Eq.~(\ref{eq:NT1}).


\begin{thebibliography}{99}
\bibitem{R-lat}
              O. Philipsen, arXiv:0710.1217




\bibitem{R-ion}
              P. Braun-Munzinger and J. Stachel, Nature {\bf 448} (2007) 302.


\bibitem{R-star}
              For recent reviews,\\              
              F. Weber, Prog. in Part. and Nucl. Phys. {\bf 54} (2005) 193.\\
              D. Page and S. Reddy, Ann. Rev. of Nucl. and
	Part. Sci. {\bf 56} (2006) 327.\\
              J.M. Lattimer and M. Prakash, Phys. Rep. {\bf 442} (2007) 109.


\bibitem{R-star2}
               T. Schaefer, arXiv:hep-ph/0509068.\\
                P. Braun-Munzinger and J. Wambach, arXiv:0801.4256. 

\bibitem{bay04} G. Baym and C.J. Pethick, {\it Landau Fermi-Liquid
	Theory} (WILEY-VCH, 2004).\\
                A.A. Abrikosov, L.P. Gorkov and I.E. Dzaloshinski, {\it
	Methods of Quantum Field Theory in Statistical Physics}
	(Prentice-Hall. Inc., 1963).\\
                A.B. Migdal, {\it Theory of finite Fermi systems}
	(Intersci. Pub., 1967).

\bibitem{noz}P. Nozi\'eres, {\it Theory of Interacting Fermi Systems}
	 (Westview Press,1997).\\
            D. Pines and P. Nozi\'eres, {\it The Theory of Quantum
	Liquids} (Perseus books Pub., 1999).

\bibitem{bay76} G. Baym and S.A. Chin, Nucl. Phys. {\bf A262} (1976)
	527.

\bibitem{tat00} T. Tatsumi, Phys. lett. {\bf B489} (2000) 280.\\
                T. Tatsumi, E. Nakano and K. Nawa, {\it Dark
	Matter}, p.39 (Nova Science Pub., New York, 2006).

\bibitem{nak03} E. Nakano, T. Maruyama and T. Tatsumi, Phys. Rev. {\bf
	D68} (2003) 105001.\\
                T. Tatsumi, E. Nakano and T. Maruyama,
	Prog. Theor. Phys. Suppl. {\bf 153} (2004) 190.\\
                T. Tatsumi, T. Maruyama and E. Nakano, {\it Superdense
	QCD Matter and Compact Stars}, p.241 (Springer, 2006).

\bibitem{nie05} A. Niegawa, Prog. Theor. Phys. {\bf 113} (2005) 581.

\bibitem{mag}
                P.M. Woods and C. Thompson, {\it Soft gamma ray
	repeaters and anomalous X-ray pulsars:magneter candidates},
	Compact stellar X-ray sources, 2006, 547.\\
                A.K. Harding and D. Lai, Rep. Prog. Phys. {\bf 69}
	(2006) 2631.

\bibitem{blo29} F. Bloch, Z. Phys. {\bf 57} (1929) 545;

\bibitem{her66} C. Herring, {\it Exchange Interactions among Itinerant
	Electrons: Magnetism IV} (Academic press, New
	York, 1966)\\
               K. Yoshida, {\it Theory of magnetism} (Springer, Berlin, 1998).

\bibitem{mor85} T. Moriya, {\it Spin Fluctuations in Itinerant Electron
Magnetism} (Springer, 1985)

\bibitem{hon} D.K. Hong, Phys. Lett. {\bf B473} (2000) 118;
	Nucl. Phys. {\bf B582} (2000) 451.

\bibitem{sch} T. Sch\"afer and K. Schwenzer, Phys. Rev. {\bf D70} (2004) 054007; 114037.

\bibitem{han04} S. Hands, Phys. Rev. {\bf D69} (2004) 014020.

\bibitem{kap} J.I. Kapusta, {\it Finite-temperature field
	theory} (Cambridge U. Press, 1993).\\
                M. Le Bellac, {\it Thermal Field Theory} (Cambridge
	U. Press, 1996).

\bibitem{tat08} T. Tatsumi, {\it Exotic States of Nuclear Matter} (World
	Sci., 2008) 272.

\bibitem{tat082} T. Tatsumi and K. Sato, Phys. Lett. {\bf B663} (2008) 322.

\bibitem{bru57} K.A. Brueckner and K. Sawada, Phys. Rev. {\bf 112}
(1957) 328.\\
         B.S. Shastray, Phys. Rev. Lett. {\bf 38} (1977) 449.

\bibitem{sha94} R. Shanker, Rev. Mod. Phys. {\bf 66} (1994) 129.

\bibitem{nay} C. Nayak and F. Wilczek, Nucl. Phys. {\bf B430} (1994)
	534.\\
J. Polchinski, Nucl. Phys. {\bf B422} (1994) 617.

\bibitem{boy01} D. Boyanovsky and H.J. de Vega, Phys. Rev. {\bf D63}
	(2001) 034016; 114028.

\bibitem{man}C. Manuel and Le Bellac, Phys. Rev. {\bf D55}  (1997) 3215.\\
             C. Manuel, Phys. Rev. {\bf D62} (2000) 076009.

\bibitem{lut} J.M. Luttinger and J.C. Ward, Phys. Rev. {\bf 118} (1960)
	1417.\\
J.M. Luttinger, Phys. Rev. {\bf 119} (1960) 1153.

\bibitem{lut61} J.M. Luttinger, Phys. Rev. {\bf 121} (1961) 942.


\bibitem{hol} T. Holstein, R.E. Norton and P. Pincus, Phys. Rev. {\bf
	B8} (1973) 2649.

\bibitem{rei} M.Yu. Reizer, Phys. Rev. {\bf B40} (1989) 11572; {\bf B44}
	(1991) 5476.

\bibitem{gan} J. Gan and E. Wong, Phys. Rev. Lett. {\bf 71} (1993) 4226.

\bibitem{cha} S. Chakravarty, R.E. Norton and O.F. Syljuasen, Phys. Rev. Lett. {\bf
	74} (1995) 1423.


\bibitem{ipp} A. Ipp, A. Gerhold and A. Rebhan, Phys. Rev. {\bf D69}
	(2004) 011901.\\
              A. Gerhold, A. Ipp and A. Rebhan, Phys. Rev, {\bf D70} (2004) 105015; PoS (JHW2005) 013.


\bibitem{ger05} A. Gerhold and A. Rebhan, Phys. Rev. {\bf D71} (2005) 085010.

\bibitem{son} D.T. Son, Phys. Rev. {\bf D59} (1999) 094019.

\bibitem{sch99}T. Sch\"afer and F. Wilczek, Phys. Rev. {\bf D60} (1999)
	114033.

\bibitem{bro} W.E. Brown, J.T. Liu and H. Ren, Phys. Rev. {\bf D61}
	(2000) 114012; {\bf D62}(2000) 054013.

\bibitem{wan} Q. Wang and D.H. Rischke, Phys. Rev. {\bf D65} (2002) 054005.

\bibitem{tat083} T. Tatsumi and K. Sato, Phys. Lett. {\bf B672} (2009) 132.

\bibitem{itz} C. Itzykson and J.-B. Zuber, {\it Quantum Field
Theory} (McGraw-Hill Inc., 1980). 

\bibitem{ber} V.B. Berestetsii, E.M. Lifshitz and L.P. Pitaevsii,\\ 
{\it Relativistic Quantum Theory} (Pergamon Press, 1971).

\bibitem{ohn} K. Ohnishi, M. Oka and S. Yasui, Phys. Rev. {\bf D76}
	(2007) 097501.

\bibitem{mit}
             T.A. DeGrand, R.L. Jaffe, K. Johnson and J. Kiskis,
	Phys. Rev. {\bf D12} (1975) 2060.

\bibitem{inu} M. Inui, H. Kohyama and A. Niegawa, arXiv:0709.2204.

\bibitem{pal} K. Pal, S. Biswas and A.K. Dutt-Mazumder, arXiv:0809.0404.

\bibitem{nam}
             S.-il Nam, H.-Y. Ryu, M.M. Musakhanov and H.-C. Kim, arXiv:0804.0056.

\bibitem{csc}
            For a recent review, M.G. Alford, K. Rajagopal, T. Schaefer, A. Schmit, arXiv:0709.4635.

\bibitem{uni}
            D. Boyanovsky, H.J. de Vega, D.J. Schwarz,
	Ann. Rev. Nucl. Part. Sci. (2006) 441.


\end{thebibliography}
\end{document}